\documentclass[aps,amssymb,eqsecnum,showkeys,showpacs]{revtex4}
\usepackage{graphicx}
\usepackage{subfigure}
\usepackage{mathptmx}
\usepackage{psfig,epsfig}

\newcommand{\be}{\begin{eqnarray}}
\newcommand{\ee}{\end{eqnarray}}
\newcommand{\bx}{{\bf x}}
\newcommand{\by}{{\bf y}}
\newcommand{\bz}{{\bf z}}
\newcommand{\br}{{\hat {\bf r}}}
\newcommand{\bs}{{\hat {\bf s}}}
\newcommand{\bt}{{\hat {\bf t}}}
\begin{document}


\title{A study of  $q {\bar q}$ states in transverse 
lattice QCD using alternative
fermion formulations}
\author{Dipankar Chakrabarti}
\email{dipankar@theory.saha.ernet.in}
\author{A. Harindranath}
\email{hari@theory.saha.ernet.in}
\affiliation{
Theory Group, Saha Institute of Nuclear Physics,
1/AF, Bidhannagar, Kolkata 700064 India}
\author{James P. Vary}
\email{jvary@iastate.edu}
\affiliation{Department of Physics and Astronomy, Iowa
State University, Ames, IA 50011, U.S.A.} 

\date{September 29,  2003}
\begin{abstract}

In this work we investigate $q{\bar q}$ spectra and wavefunctions of 
 light front transverse lattice
Hamiltonians that result from different methods of formulating fermions on the
transverse lattice. We adopt the one link approximation for the transverse
lattice and Discrete Light Cone Quantization to handle longitudinal dynamics. 
We perform a detailed study of the continuum limit of DLCQ and
associated techniques to
manage severe light front infrared divergences. We explore the effects
 of various parameters of the
theory, especially, the strength of the helicity-flip interaction and the link mass
on spectra and wavefunctions. 

\end{abstract}
\pacs{11.10.Ef, 11.15.Ha, 11.15.Tk, 12.38.-t}
\keywords{transverse lattice, fermions, $q{\bar q}$ states}
\maketitle
\section{Introduction}
 A promising method to calculate observables in QCD is the transverse
lattice formulation \cite{Bardeen:1976tm,Bardeen:1980xx,review}. 
In this method, 
one keeps $x^\pm = x^0 \pm x^3$ continuous and discretizes the transverse
space spanned by coordinates $x^\perp =(x^1, x^2)$. With the gauge choice
$A^+=0$, $A^-$ becomes a constrained variable which can be eliminated in
favor of dynamical gauge variables. So far very encouraging results have
been obtained in the pure gauge and meson sectors \cite{dalme,buseal,vandaly}. 

 Due to the doubling phenomena, fermions on the lattice pose challenging
problems. To date, calculations of meson properties using transverse lattice
have employed Wilson fermions \cite{BK}. It is well-known that the Wilson term
explicitly breaks the chiral symmetry and makes it difficult to explore the
consequences of spontaneous chiral symmetry breaking in the chiral limit. In
this limit, in one link approximation 
on the transverse lattice, the Wilson term 
can be adjusted to
produce the desirable level splitting between $ \pi$ and $ \rho$. This,
however, results in the undesirable consequence that the 
splitting of the $ \rho$
multiplet is almost as large as the $ \pi - \rho$ splitting.  Because of the
doubling problem, one cannot keep the Wilson term very small. Thus it is
desirable to explore other formulations of fermions on the transverse
lattice that may have different chiral properties. 

In a recent work \cite{Chakrabarti:2002yu} 
we have addressed the problems of
fermions on the light front
transverse lattice. We proposed and numerically investigated different
approaches of formulating fermions on the transverse lattice. In one
approach, which uses forward and backward derivatives, fermion doubling is
absent and the helicity flip term which is proportional to fermion mass in
light front QCD becomes an irrelevant term in the free field limit. In the
literature, symmetric derivatives have been used which leads to fermion
doubling due to the decoupling of even and odd lattices. Using the light front
staggered fermion formulation and the Wilson fermion formulation, we
studied the removal of doublers from the spectrum. Our investigations lead
to the identification of an even-odd helicity flip symmetry of the light
front transverse Hamiltonian, the absence of which means the removal of doublers in
all the cases that we studied. 

In this work we make a detailed comparison of 
various light front QCD 
Hamiltonians that result from different ways of formulating fermions on the
transverse lattice. As the first step in our calculations, we adopt
the one link approximation in the meson sector which has been widely used in
the literature. (Only very recently, the effect of additional links in the
meson sector has been investigated \cite{vandaly}). 
Since the one link
approximation is very crude, our aim is not to reproduce physical
observables. Rather, we explore the effects of various coupling strengths on
the low-lying spectra and wave functions and compare two different formulations. 

We use Discretized Light Cone Quantization (DLCQ) \cite{dlcq} 
to address 
longitudinal
dynamics. Because of the presence of severe light front infrared
divergences, a major concern here is the reliability of DLCQ results when
calculations are done at finite resolution $K$ and results are extrapolated
to the continuum  ($ K \rightarrow \infty$). In meson calculations so far, 
$K \leq 20$ have been chosen. In this work we perform a detailed study of
the continuum limit of DLCQ by performing 
calculations at larger values of
$K$.   

In the meson sector, in the zero link approximation, at each transverse
location we have a two-dimensional field theory which in the large $N_c$ limit
(where $N_c$ is the number of colors) is
nothing but the 't Hooft model. In this well-studied model, excited states
are simply excitations of the $q {\bar q}$ pair, which contain nodes in
the wavefunctions. The picture changes when one link is included thereby
allowing fermions to hop. The admixture of $q {\bar q}$ link states with $q
{\bar q}$ states is controlled by the strengths of the particle number
changing 
interactions and the mass of the link field. One link approximation is a
priori justified for very massive links and/or weak particle changing
interaction since in this case low lying excited states are also $q{\bar q}$
excitations. Likewise, for large particle changing 
interaction strength and/or light
link mass, low lying excited states are $q{\bar q}$ link states. We explore the
spectra and wavefunctions resulting from the choice of various regions of
parameter space.

The plan of this paper is as follows. In Sec. II we 
present the details of the 
light front transverse lattice Hamiltonian resulting from the use of
forward and backward derivatives and the resulting effective Hamiltonian
when the unitary link variables are replaced by general complex matrices.
In this section we also present the canonical transverse lattice QCD 
Hamiltonian resulting from the addition of the Wilson term.
Sec. III contains comparisons of numerical results for the two
Hamiltonians.
Finally Sec. IV contains our summary and conclusions.
Typical terms in the Hamiltonian with forward and backward derivatives in
the Fock representation in DLCQ is presented in Appendix \ref{appa}. Explicit
expressions for the states are given in Appendix \ref{appb}.
For completeness, explicit
expressions for the matrix elements  in the 
forward-backward case and the  
Wilson case are presented in Appendices \ref{appc} and \ref{appd}.
\section{Hamiltonians} 
Due to the constraint equation in light front theory, different
methods are possible to put fermions on the transverse lattice. In this
section we present the detailed structure of the resulting QCD Hamiltonians 
for two methods studied in Ref.
\cite{Chakrabarti:2002yu}, namely, forward and backward derivatives and
symmetric derivatives together with Wilson term.  

\subsection{Hamiltonian with forward and backward derivatives}
Details of the derivation of the fermionic part of the 
Hamiltonian are already 
given in  Ref. \cite{Chakrabarti:2002yu}. Here we give the details of the
gauge field part of the Hamiltonian. Non-linear constraints on the unitary
link variables make it difficult to perform canonical quantization. We also
present the effective Hamiltonian when non-linear unitary variables are
replaced by linear variables. 
\subsubsection{Gauge field part of the Lagrangian density}
 
The gauge field part of the Lagrangian density in the continuum is
\be
{ {\cal L}}_G = { 1 \over 2 g^2} Tr F_{\rho \sigma}F^{\rho \sigma}
\ee
where $ F^{\rho \sigma} = \partial^\rho A^\sigma - \partial^\sigma
A^\rho +  ~[ A^\rho, A^\sigma]$ with $ A^\rho = i g A^{\rho \alpha} T^\alpha$.
Here $ \rho, \sigma=0,1,2,3$ and $ \alpha=1,2, \ldots, 8.$ For ease of
notation we suppress the dependence of field variables on the longitudinal
coordinate in this section.
With the gauge choice $A^+=0$,  the Lagrangian density can be separated into
three parts,
\be
{\cal L}_G = {\cal L}_T + {\cal L}_L + {\cal L}_{LT}.
\ee
Here $ {\cal L}_T $ depends entirely on the lattice gauge field $U_r(\bx)$.
\be
{\cal L}_T = { 1 \over g^2 a^4 } \sum_{r \neq s} \Bigg \{
Tr \Big [
U_r(\bx) U_s(\bx + a \br) U_{-r}(\bx + a \br + a \bs)U_{-s}(\bx + a
\bs) 
 -1 \Big ] \Bigg \}.
\ee
The  purely longitudinal part $ {\cal L}_{L}$ depends on the constrained
gauge field $A^-$, 
\be
{\cal L}_{L}= { 1 \over 8} (\partial^+ A^{- \alpha})^2
\ee
and the mixed part $ {\cal L}_{LT}$ depends both on lattice gauge field and the
constrained gauge field. 
\be
{\cal L}_{LT} = {1 \over g^2 a^2} Tr [ \partial_\mu U_r(\bx)
\partial^\mu U_r^\dagger(\bx)] + {1 \over 2 a^2}g A^{-\alpha}J^{+\alpha}_{LINK}.
\ee
Here the link current
\be J^{+\alpha}_{LINK} (\bx)= \sum_r {1 \over g^2}Tr \Bigg \{ T^\alpha 
[  U_r(\bx) i\stackrel{\leftrightarrow}{\partial^+} U_r^\dagger (\bx)+  
U_r^\dagger (\bx- a \br) 
i \stackrel{\leftrightarrow}{\partial^+} U_r (\bx - a \br)] \Bigg \}.
\ee

Substituting back the expression for $A^{-\alpha}$ from the constraint
equation
\be
(\partial^+)^2 A^{-\alpha} = \frac{2g}{a^2} \left (J^{+\alpha}_{LINK} - 
J^{+\alpha}_q) \right )
\ee
with 
\be J^{+\alpha}_q (\bx)= 2 \eta^\dagger(\bx) T^\alpha \eta(\bx) 
\ee
where $\eta$ is the dimensionless two-component lattice fermion field,
in the $A^{-\alpha}$ dependent terms in the Lagrangian density, namely,
\be
- { 1 \over 2} \frac{g}{a^2} A^{-\alpha} J^{+ \alpha}_q + 
{ 1 \over 8}  (\partial^+ A^{- \alpha})^2  + { 1 \over 2} \frac{g}{a^2}
A^{- \alpha}J^{+\alpha}_{LINK}
\ee
we generate the terms
\be
{g^2 \over 2 a^4} J^{+\alpha}_{LINK}\left ({ 1\over \partial^+}\right)^2
J^{+\alpha}_{LINK}+ {g^2 \over 2 a^4} \eta^\dagger T^\alpha  \eta   
\left ({ 1\over \partial^+}\right )^2  \eta^\dagger T^\alpha  \eta -
 { g^2 \over a^4} 
J^{+\alpha}_{LINK} \left ({ 1\over \partial^+}\right)^2 \eta^\dagger T^\alpha  \eta.
\ee
Collecting all the terms, the canonical Lagrangian density for
transverse lattice QCD is
\be
{\cal L}  &=& {1\over a^2}\eta^\dagger (\bx) i \partial^- \eta (\bx) 
+{ 1 \over a^4 g^2} Tr [ \partial_\mu U_r(\bx)
\partial^\mu U_r^\dagger(\bx)]
- {m^2\over a^2}
\eta^\dagger (\bx) { 1 \over i \partial^+} \eta(\bx) \nonumber \\ 
&~&  + im{1\over a^2} \eta^\dagger(\bx)  {\hat
\sigma}_s {1 \over a}{1 \over \partial^+}\Big [
U_s(\bx) \eta(\bx + a \bs) - \eta(\bx) \Big ] 
\nonumber \\
&~& +im {1\over a^2}
\Big [
\eta^\dagger(\bx + a \br) U_r^\dagger(\bx) -
\eta^\dagger(\bx) \Big ] 
 {\hat \sigma}_r{ 1 \over a} { 1 \over 
\partial^+} \eta(\bx) \nonumber \\
&~& - { 1 \over a^4} [ \eta^\dagger(\bx + a \br) U_r^\dagger(\bx) -
\eta^\dagger(\bx)] {\hat \sigma}_r { 1 \over i \partial^+}{\hat
\sigma}_s [ U_s(\bx) \eta(\bx + a \bs) - \eta(\bx)] \nonumber \\
&~& +  {1 \over a^4 g^2} \sum_{r \neq s}
\Bigg \{ Tr \Big [
U_r(\bx) U_s(\bx + a \br) U_{-r}(\bx + a \br + a \bs)U_{-s}(\bx + a
\bs) 
- 1\Big ] \Bigg \} \nonumber \\
&~& + {g^2 \over 2 a^4} J^{+\alpha}_{LINK}\left({ 1\over \partial^+}\right)^2
J^{+ \alpha}_{LINK}+ {1 \over 2 a^4} g^2 J^{+\alpha}_q   
\left ({ 1\over \partial^+}\right )^2  J^{+\alpha}_q \nonumber \\
&~&~~~~ - { g^2 \over a^4} 
J^{+\alpha}_{LINK} \left ({ 1\over \partial^+}\right )^2 
J^{+\alpha}_q. \label{fb_L}
\ee
Here ${\hat \sigma}_1=\sigma_2, ~{\hat \sigma}_2 =-\sigma_1$.
In the two-component representation \cite{hz}, the dynamical fermion field
\be
\psi^+(x^-, x^{\perp}) = \left[ \begin{array}{l} {1 \over a}\eta(x^-, x^{\perp})
\\ 0 \end{array} \right] 
\ee
where $\eta$ is the dimensionless two component lattice fermion field.

\subsubsection{Effective Hamiltonian}
Because of the nonlinear constraints $ U^\dagger U= 1$, $ det
~U=1$, 
it is
highly nontrivial to quantize the system. Hence 
Bardeen and Pearson \cite{Bardeen:1976tm}
and Bardeen, Pearson, and Rabinovici \cite{Bardeen:1980xx} proposed to 
replace the
nonlinear variables $U$ by linear variables $M$ where $M$ belongs to
$GL(N,{\cal C})$, i.e., we replace 
$ { 1 \over g} U_r(\bx) \rightarrow M_r(\bx)$.  Once we replace $U$ 
by $M$, many more terms are
allowed in the Hamiltonian. Thus one needs to add an effective potential
$V_{eff}$ to the Lagrangian density
\be
V_{eff}=  - {\mu^2\over a^2} ~Tr (M^\dagger M) + \lambda_1 ~Tr[(M^\dagger M)^2] +
\lambda_2 ~[det~ M + H.c] + \ldots .
\ee

Thus, the effective Hamiltonian for QCD on the transverse
lattice becomes
\be
P^-_{fb} & = & P^-_{f~free} + P^-_V+ P^-_{fhf}+P^-_{hf} 
+ P^-_{chnf} \nonumber \\
&~& + P^-_{qqc} + P^-_{ggc} + P^-_{qgc} + P^-_{p}.
\ee
The free fermion part is
\be  P^-_{f~free} & = & \int dx^-  \sum_{\bx} (m^2 + {2 \over a^2})
\eta^\dagger (\bx) { 1 \over i \partial^+} \eta(\bx)
\ee
The effective potential part is 
\be 
 P^-_V & = & 
\int dx^- a^2 \sum_\bx \Bigg ({ \mu^2\over a^2} ~Tr (M^\dagger M)
- \lambda_1 ~Tr[(M^\dagger M)^2] -
\lambda_2 ~[det~ M + H.c] + \ldots  \Bigg ). \nonumber \\
\ee
The free helicity-flip part is 
\be 
P^-_{fhf} & = & 2im\int dx^-\sum_{\bx}\sum_s ~\eta^\dagger(\bx)  {\hat
\sigma}_s {1 \over a}{1 \over \partial^+} \eta(\bx). 
\ee
Helicity flip associated with the fermion hop is
\be
P^-_{hf} & = & - img \int dx^-\sum_{\bx} \sum_s \eta^\dagger(\bx)  {\hat
\sigma}_s {1 \over a}{1 \over \partial^+}\Big [
M_s(\bx) \eta(\bx + a \bs) \Big ]
\nonumber \\
&~& -img \int dx^-\sum_{\bx}\sum_r
\Big [
\eta^\dagger(\bx + a \br)  ~M_r^\dagger(\bx) \Big ]
 {\hat \sigma}_r{ 1 \over a} { 1 \over
\partial^+} \eta(\bx).
\ee
Canonical helicity non-flip terms are
\be
P^-_{chnf} & = & - { g \over a^4}\int dx^- a^2\sum_{\bx} \sum_{rs} 
[ \eta^\dagger(\bx + a \br)  M_r^\dagger(\bx) ] 
{\hat \sigma}_r { 1 \over i \partial^+}{\hat
\sigma}_s [  \eta(\bx)] \nonumber \\
&~& - { g \over a^4}\int dx^- a^2\sum_{\bx} \sum_{rs} 
[\eta^\dagger(\bx)] {\hat \sigma}_r { 1 \over i \partial^+}{\hat
\sigma}_s [  M_s(\bx) \eta(\bx + a \bs)] \nonumber \\
&~& - { g^2 \over a^4} \int dx^- a^2\sum_{\bx}\sum_{rs} [ \eta^\dagger(\bx + a \br)
~M_r^\dagger(\bx)
] {\hat \sigma}_r { 1 \over i \partial^+}{\hat
\sigma}_s [  M_s(\bx) \eta(\bx + a \bs) ]. \label{canhnf}
\ee
The four-fermion instantaneous term is
\be
P^-_{qqc} & = & -2 {g^2\over a^2} \int dx^-  \sum_\bx
 \eta^\dagger(\bx) T^a \eta(\bx) { 1 \over (\partial^+)^2}
\eta^\dagger(\bx) T^a \eta(\bx) .
\ee
The four link instantaneous  term is
\be
P^-_{ggc} & = & - \frac{1}{2} {g^2 \over a^2}\int dx^-  \sum_\bx
  J^{+a}_{LINK}(\bx) { 1 \over
(\partial^+)^2}J^{+a}_{LINK}(\bx). 
\ee
The fermion - link instantaneous term is
\be
P^-_{qgc} & = &  2 {g^2\over a^2}\int dx^- \sum_\bx
 J^{+a}_{LINK}(\bx) { 1 \over (\partial^+)^2}
\eta^\dagger(\bx)
T^a \eta(\bx),
\ee
The plaquette term is
\be 
P^-_{p} & = & -  {g^2 \over a^4} \int dx^- a^2\sum_\bx \sum_{r \neq s}
\Bigg \{ Tr \Big [
M_r(\bx) M_s(\bx + a \br) M_{-r}(\bx + a \br + a \bs)M_{-s}(\bx + a \bs)
- 1\Big ] \Bigg \}.
\ee
Here
\be 
J^{+\alpha}_{LINK} (\bx)= \sum_r Tr \Bigg \{ T^\alpha
[  M_r(\bx) i \stackrel{\leftrightarrow}{ \partial^+} M_r^\dagger (\bx)+
M_r^\dagger (\bx- a \br)
i \stackrel{\leftrightarrow}{ \partial^+} M_r (\bx - a \br)] \Bigg \}.
\ee

\subsubsection{Violations of hypercubic symmetry}
The canonical helicity non-flip interactions given in Eq.
(\ref{canhnf}) for $ r \neq s$  break the  hypercubic symmetry on the transverse lattice. 
For interacting theory this is also true for the  Hamiltonian with 
symmetric derivative.  In the free field limit they do not survive for
 Hamiltonian with symmetric derivative but for forward-backward derivative 
they survive. In that case, 
in the free field limit they reduce to 
\be
\frac {1}{a^2}~\int dx^- \sum_{\bx}\sum_{r \neq s}\Bigg [ 
\eta^\dagger(\bx + a \br) {\hat \sigma}_r {\hat \sigma}_s \frac{1}{
\partial^+} \eta(\bx) \nonumber \\
+ \eta^\dagger(\bx) {\hat \sigma}_r {\hat \sigma}_s \frac{1}{
\partial^+} \eta(\bx + a \bs) \nonumber \\
- \eta^\dagger(\bx + a \br) {\hat \sigma}_r {\hat \sigma}_s 
\frac{1}{
\partial^+} \eta(\bx + a \bs) \Bigg ].
\ee
Going to the transverse momentum space via
\be
\eta(x^-, x^\perp) = \int d^2 k^\perp e^{i k^\perp \cdot x^\perp} ~
\phi_{k^{\perp}}(x^-)
\ee
we get
\be
- \frac{2}{a^2} \int dx^- \int d^2 k^\perp  \phi^\dagger _{k^{\perp}}(x^-)
\sigma_3 \frac{1}{ i \partial^+} \phi_{k^{\perp}}(x^-) \nonumber \\
\Big [ \sin (k_ya)  - \sin (k_xa) + \sin (k_x a - k_y a) \Big ]. 
\ee
Thus the violations of hypercubic symmetry are of the order of the lattice
spacing $a$. Sign in front of this term changes if we switch forward and  
backward derivatives.
  
In our numerical studies presented in Ref. \cite{Chakrabarti:2002yu} 
and in this work, 
we have set the coefficients of hypercubic symmetry violating terms to zero. 
\subsection{Canonical transverse lattice QCD with the Wilson term}
When one uses symmetric derivatives for the fermion fields,  
doublers arise as a result of the decoupling of even and odd lattice sites. To
remove the doublers one may use the Wilson fermions \cite{BK} or the
Kogut-Susskind fermions \cite{griffin,Chakrabarti:2002yu}. In this
subsection, the details of the structure of the Hamiltonian resulting with the
modification of the Wilson term are presented.  

\subsubsection{Constraint equation}
The symmetric derivative is defined by
\be
D_r \psi^\pm(\bx) = { 1 \over 2 a} [ U_r(\bx) \psi^\pm(\bx + a \br) - 
U_{-r}(\bx) \psi^\pm(\bx - a \br)].
\ee
Again we make the replacement $ \frac{1}{g} U_r(\bx) = M_r(\bx)$.
Anticipating doublers, we can add a $``$Wilson term'':
\be
\delta {\cal L}&=& \frac{\kappa}{a}{\bar \psi}(\bx) [g M_r(\bx)
\psi(\bx + a \br) - 2 \psi(\bx) + g M_{-r}(\bx) \psi(\bx - a \br)]~
\ee
where $ \kappa$ is the dimensionless Wilson parameter. Explicitly, in terms of the
dynamical field $\psi^+$
and the constrained $ \psi^-$
\be
\delta {\cal L}&=& { \kappa \over a} {\psi^-}^\dagger (\bx) \gamma^0 
[g M_r(\bx) \psi^+(\bx + a \br) - 2 \psi^+(\bx) + g M_{-r}(\bx)
\psi^+(\bx - a \br)] \nonumber \\
&~& +  { \kappa \over a} {\psi^+}^\dagger (\bx) \gamma^0 
[g M_r(\bx) \psi^-(\bx + a \br) - 2 \psi^-(\bx) + gM_{-r}(\bx)
\psi^-(\bx - a \br)].
\ee
The constraint equation for $\psi^-$ in the presence of the Wilson term is
\be
i \partial^+ \psi^-(\bx) & = & m \gamma^o \psi^+(\bx) \nonumber \\
&~& + i {\alpha_r \over 2 a} [ gM_r(\bx) \psi^+(\bx + a \br) -
gM_{-r}(\bx) \psi^+(\bx - a \br)] \nonumber \\
&~& - { \kappa \over a} \gamma^0 [ gM_r(\bx) \psi^+(\bx + a \br)- 2
\psi^+(\bx)
+ gM_{-r}(\bx) \psi^+(\bx - a \br)].
\ee

\subsubsection{Hamiltonian: Symmetric derivatives and the
Wilson term}

After a great deal of algebra, we arrive at the Hamiltonian,
\be
 P^- & = & P^-_{f~free} + P^-_V+ P^-_{hf} + P^-_{whf}  \nonumber \\
&~& +P^-_{chnf} + P^-_{wnf1} + P^-_{wnf2} \nonumber \\
&~& + P^-_{qqc} + P^-_{ggc} + P^-_{qgc} + P^-_{p}.
\ee
The free fermion part is
\be
P^-_{f~free}= \int dx^- a^2 \sum_\bx
{1\over a^2}\left(m+4 \frac{\kappa}{a}\right)^2
\eta^\dagger(x^-, \bx) { 1 \over i \partial^+} \eta(x^-, \bx) .
\ee
The helicity flip part is
\be
P^-_{hf} & = &  -g \int dx^-  \sum_\bx
\Bigg \{ \left ( m+ 4 \frac{\kappa}{a} \right ) \frac{1}{2a}
\eta^\dagger(\bx) \sum_r {\hat \sigma}_r { 1 \over i \partial^+}  
\left [ M_{r}(\bx) \eta(\bx + a \br) - M_{-r}(\bx) \eta(\bx - a \br) \right
] 
\nonumber \\
& ~ & - \left ( m+ 4 \frac{\kappa}{a} \right ) \frac{1}{2a}
 \sum_r \left [ \eta^\dagger(\bx - a \br) {\hat \sigma}_rM_{r}(\bx - a \br)
- 
 \eta^\dagger(\bx + a \br) {\hat \sigma}_r M_{-r}(\bx + a \br)  \right ]
{ 1 \over i \partial^+} \eta(\bx) \Bigg \} . \nonumber \\
\ee
The Wilson term induced helicity flip part
\be
P^-_{whf} & = & g^2\int dx^-  \sum_\bx \Bigg \{
 \frac{\kappa}{a} \frac{1}{2a}\sum_r \sum_s 
\Big [ \eta^\dagger(\bx - a \br) M_{r}(\bx - a \br) + 
\eta^\dagger(\bx + a \br) M_{-r}(\bx + a \br) \Big ] \nonumber \\
&~&~~~~~{ 1 \over i \partial^+} {\hat \sigma}_s 
\Big [  M_{s}(\bx) \eta(\bx + a \bs) - M_{-s}(\bx) \eta(\bx - a \bs) \Big ]
\nonumber \\
& ~ & - \frac{\kappa}{a} \frac{1}{2a}\sum_r \sum_s 
\Big [ \eta^\dagger(\bx - a \br) {\hat \sigma}_r M_{r}(\bx - a \br) - 
\eta^\dagger(\bx + a \br) {\hat \sigma}_r M_{-r}(\bx + a \br) \Big ]
\nonumber \\
&~&~~~~~{ 1 \over i \partial^+}  
\Big [  M_{s}(\bx) \eta(\bx + a \bs) + M_{-s}(\bx) \eta(\bx - a \bs)\Big ]
\Bigg \}.
\ee
The canonical helicity non-flip term arising from fermion constraint is 
\be
P^-_{chnf} & = & -g^2 \int dx^-  \sum_\bx
 \frac{1}{4a^2} \sum_r \sum_s 
\Big [ \eta^\dagger(\bx - a \br) {\hat \sigma}_r M_{r}(\bx - a \br) - 
\eta^\dagger(\bx + a \br)  {\hat \sigma}_r M_{-r}(\bx + a \br) \Big ]
\nonumber \\
&~&~~~~~{ 1 \over i \partial^+} {\hat \sigma}_s 
\Big [  M_{s}(\bx) \eta(\bx + a \bs) - M_{-s}(\bx) \eta(\bx - a \bs)\Big ] .
\ee
The Wilson term induced helicity non flip terms are  
\be
P^-_{wnf1} & = & -g \int dx^-  \sum_\bx
\Bigg \{\left ( m+ 4 \frac{\kappa}{a} \right ) \frac{\kappa}{a}
\eta^\dagger(\bx) { 1 \over i \partial^+}\sum_r    
\left [ M_{r}(\bx) \eta(\bx + a \br) 
+ M_{-r}(\bx) \eta(\bx - a \br) \right ] \nonumber \\
& ~ & + \left ( m+ 4 \frac{\kappa}{a} \right ) \frac{\kappa}{a}
 \sum_r \left [ \eta^\dagger(\bx - a \br) M_{r}(\bx - a \br) + 
 \eta^\dagger(\bx + a \br) M_{-r}(\bx + a \br)  \right ]
{ 1 \over i \partial^+} \eta(\bx) \Bigg \}. \nonumber \\
\ee
and 
\be
P^-_{wnf2} & = & - g^2 \int dx^-  \sum_\bx
  \frac{\kappa^2}{a^2} \sum_r \sum_s 
\Big [ \eta^\dagger(\bx - a \br)  M_{r}(\bx - a \br) + 
\eta^\dagger(\bx + a \br)  M_{-r}(\bx + a \br) \Big ] \nonumber \\
&~&~~~~~{ 1 \over i \partial^+}  
\Big [  M_{s}(\bx) \eta(\bx +a \bs)  + M_{-s}(\bx) \eta(\bx - a \bs)\Big ].
\ee

Comparing the Hamiltonians with a) forward-backward derivative and b)
symmetric derivative with the Wilson term we notice that the only
differences are in the particle number changing interactions, namely, helicity flip
 and helicity non-flip terms. 
\section{One link approximation}
\subsection{Relevant interactions}
In one link approximation, for either Hamiltonian, the four link
instantaneous  term and the plaquette term do not contribute and only
the link mass term of the effective potential contributes.  Further, in the
case of the forward-backward Hamiltonian, the helicity non-flip part
proportional to $g^2$ does not contribute. For the Wilson term modified
Hamiltonian, the Wilson term induced helicity flip part $P^-_{whf}$, the
canonical helicity non-flip term $ P^-_{cnhf}$  and the term proportional to
$\kappa^2$ in the Wilson term induced helicity non-flip part do not
contribute. Thus in the case of the Wilson term 
modified Hamiltonian the entire
fermion hopping with no helicity flip arises from the Wilson term. 
\subsection{Comparison with one gluon exchange in the continuum}
It is interesting to compare the one link approximation on the transverse
lattice with the one gluon exchange approximation in the continuum. In the
latter, a major source of singularity  is the $\frac{k^\perp}{k^+}$ term in
the quark - gluon vertex where $k^\perp$ ($k^+$) is the gluon transverse
(longitudinal) momentum. This originates from the $A^-J^+_q$ interaction term in
the Hamiltonian via $\frac{1}{\partial^+}\partial^\perp \cdot A^\perp$
contribution to the constrained field $A^-$. This term gives rise to
quadratic ultraviolet divergence in the transverse plane
accompanied by linear divergence in the longitudinal direction in fermion
self energy. 
On the transverse lattice, $ \partial^+ A^- \propto \frac{1}{\partial^+}
J^+_{LINK}$ so that $ A^- J^+_q \rightarrow J^+_{LINK}
\frac{1}{(\partial^+)^2} J^+_q$. Thus a term which gives rise to severe
divergence structure in the continuum gets buried in the fermion-link
instantaneous interaction term which gives rise to a term in the gauge boson
fermion vertex in the continuum in Abelian theory. In the non-Abelian gauge
theory this gives rise to a term in the quark-gluon vertex and also to the
instantaneous quark-gluon interaction in the continuum. 

The transfer of the troublesome term from quark-gluon vertex in the continuum
theory to quark - link instantaneous interaction term in the lattice theory
has an interesting consequence. In the continuum theory, the
addition of a gluon
mass term by hand spoils the cancellation of the light front singularity between one
gluon exchange and the instantaneous  four - fermion interaction. On the
transverse lattice this cancellation is absent anyway with or without a link
mass term.
\subsection{Longitudinal dynamics and effects of transverse hopping}

We first consider the dynamics in the absence of any link. In this case,
fermions cannot hop, and at each transverse location we have (1+1)
dimensional light front QCD which reduces to the 't Hooft model in the large
$N_c$ limit. In this case quark and antiquark at the same transverse
position interact via the spin independent instantaneous interaction 
which, in the non-relativistic limit reduces to the linear potential in the 
longitudinal direction. The only parameters in the theory are the dimensionless fermion
mass $m_f = am$ and the gauge coupling $g$.   The spectrum consists of
a ground state and a tower of excited states corresponding to the
excitations of the $q {\bar q}$ pair.

Next consider the inclusion of  the $q {\bar q}$ link states. There are four
independent amplitudes corresponding to whether the quark is left,
right, above, or below the antiquark. 
With non-zero mass of the link, these states lie above the ground
state of pure quark - antiquark system. Further the  
$q$,  ${\bar q}$  and link (which are 
frozen at their transverse positions)
undergo fermion - link instantaneous interactions in the longitudinal
direction which further increases the mass of $q {\bar q}$ link states.
Now the quark or antiquark can hop via helicity flip or helicity non-flip.
Here we find a major difference between the Hamiltonians resulting from 
forward-backward derivative  and symmetric derivative. Let us first consider 
the helicity flip hopping term in the forward-backward case 
\be
 P^-_{hf} & = &  - img \int dx^- \sum_{\bx} \sum_r \Big [ \eta^\dagger(\bx)  {\hat \sigma}_r
{1 \over a}{1 \over \partial^+}\eta(\bx + a \br) + \eta^\dagger(\bx + a \br) 
 ~M_r^\dagger(\bx)
 {\hat \sigma}_r{ 1 \over a} { 1 \over
\partial^+} \eta(\bx) \Big ]. \label{fbhf}
\ee
If we consider transition from two particle to three particle state by a quark hop, then the 
first term in Eq. (\ref{fbhf}) corresponds to $\mid 2\rangle \rightarrow \mid 3 a\rangle$ and 
the second term corresponds to  $\mid 2\rangle \rightarrow \mid 3 b\rangle$. 
The helicity flip term in symmetric derivative case,
 after making some shifts in 
lattice points, can be written as 
\be
P^-_{hf} & = &  -g \left ( m+ 4 \frac{\kappa}{a} \right )\frac{1}{2a}
\int dx^- \sum_{\bx}\sum_r 
\nonumber \\
&~& \Bigg [ \Bigg \{
\eta^\dagger(\bx) {\hat \sigma}_r { 1 \over i \partial^+}  
 M_{r}(\bx) \eta(\bx + a \br) 
-\eta^\dagger(\bx ) {\hat \sigma}_rM_{r}(\bx){ 1 \over i \partial^+} \eta(\bx +a\br) 
\Bigg \} \nonumber \\
&~& -\Bigg \{ \eta^\dagger(\bx) {\hat \sigma}_r { 1 \over i \partial^+}
 M_{-r}(\bx) \eta(\bx - a \br) 
 -\eta^\dagger(\bx) {\hat \sigma}_r M_{-r}(\bx) 
{ 1 \over i \partial^+} \eta(\bx-a\br) \Bigg \} \Bigg ]. \label{symhf}
\ee

 For the Hamiltonian with symmetric derivative, a
quark or antiquark hopping accompanied by helicity flip has opposite signs
for forward and backward hops. On the other hand, hopping accompanied by
helicity non-flip have the same signs. As a result, there is no
interference between helicity flip and helicity non-flip interactions \cite{BK}. 
In the case of the Hamiltonian with forward-backward derivative,  quark or
antiquark hopping accompanied by helicity flip has the same sign for forward
and backward hops.
 As a consequence  the helicity non-flip hop can 
interfere with the helicity flip hop.
 This has immediate 
consequences for the
spectrum. In the case with symmetric derivative, in lowest order
perturbation theory, the helicity zero states mix with 
each other which causes a
splitting in their eigenvalues resulting in the singlet state lower than the
triplet state. On the other hand, helicity plus or minus one states do not mix with each
other or with helicity zero states resulting in a two fold degeneracy. 
In the case with forward and backward
derivatives all helicity states mix with each other resulting in the complete
absence of degeneracy.  

\section{Singularities, divergence and counterterms}

Since the transverse lattice serves as an  ultraviolet regulator, we need to
worry about only light front longitudinal momentum singularities.    
\subsection{Tree level}

We take all the terms in the Hamiltonian to be normal 
ordered. At tree level this leaves us with singular factors of the form $\frac{1}{(k)^2}$
in the normal ordered four fermion and fermion link instantaneous 
interactions. The singularities are removed by adding
the counterterms used in the
previous work \cite{dalme} on transverse lattice.
The explicit forms of the
counterterms are given in Appendix \ref{appc} in the appropriate places.   
\subsection{Self energy corrections}
In the one link approximation, a quark  can make a  forward (backward) hop
followed by a backward (forward) hop resulting in self energy corrections.
In a single hop, helicity flip or non-flip can occur. In the case of
symmetric derivatives, helicity flip cannot interfere 
with helicity non-flip,
and as a consequence, self energy corrections are diagonal in helicity space.
In the case of forward and backward derivatives, the
 interference is nonzero
resulting in self energy corrections, both diagonal and 
off-diagonal in the
helicity space. Similar self energy corrections are generated for an
antiquark
also. These self energy corrections contain a 
logarithmic light
front infrared divergence which must be removed by  counterterms. In 
Appendix \ref{appcounter} we present the explicit form of counterterms in 
the two cases
separately.  In previous works on one link approximation \cite{dalme,buseal,BK},
these counterterms were not implemented.

\section{Numerical Results}
 We diagonalize the dimensionless matrix $a^2 P^-$. We further
divide the matrix elements by $g^2 C_f$ which is the 
strength of the matrix elements
for four fermion and fermion - link instantaneous interactions.
Now,  define the
constant $G$ with dimension of mass  by  $ G^2 =\frac{g^2}{a^2} C_f$.
 DLCQ yields $M^2
/G^2$.

 The dimensionless couplings are introduced \cite{dalme} as
follows. Fermion mass $ m_f=  m/G $, link  mass $ \mu_b =  \mu/G $, 
particle number conserving helicity flip coupling 
$ m_f/(aG)=m_f C_1 $, particle number non-conserving helicity flip 
$ \sqrt{N} g m_f/(aG) = m_f C_2 $, and particle number non-conserving helicity non-flip
$\sqrt{N}g /(a^2 G^2) =C_3 $. In the case of the
Wilson term modified Hamiltonian,
we have fermion mass term $ m_f = (m+ 4 \kappa/a)/G $, helicity-flip
coupling $  \sqrt{N}g m_f /(2aG) = m_f{\tilde C}_2$, and
helicity non-flip
coupling $ \sqrt{N}g m_f \kappa /(aG) =m_f {\tilde C}_3 $. 

All the results presented here were obtained on a small cluster of computers using the
 Many Fermion Dynamics (MFD) code \cite{mfd} implementing the Lanczos diagonalization 
mathod in parallel 
environment. For low K values, the  results were checked against an independent code 
running on a single processor.

\subsection{Cancellation of divergences}
\begin{figure}
\centering
\includegraphics[width=6.0in]{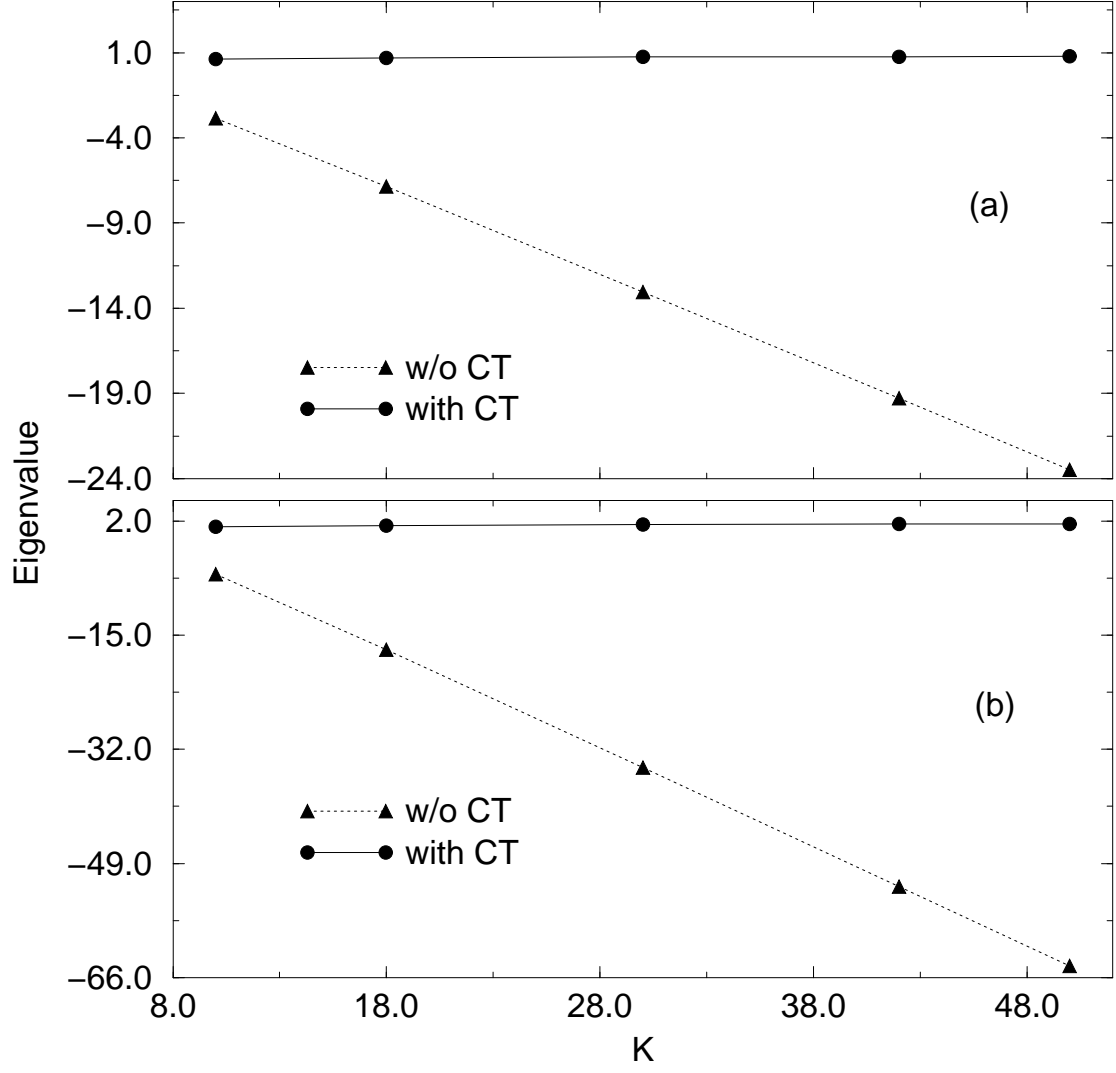}
\caption{ Effect of counterterm on the ground state eigenvalue. 
(a) With and without the counterterm in the $ q {\bar q}$ sector for $m_f=0.3$. (b) With
and without the counterterm in the $ q {\bar q}$ link sector for $m_f=0.3$ and $\mu_b=0.2$. } 
\label{CT}
\end{figure}

As we already mentioned, we encounter $\frac{1}{(k^+)^2}$ singularities 
with instantaneous four fermion and instantaneous fermion - link interactions
which give rise to linear divergences.
We remove the divergences by adding appropriately chosen counterterms.
We have numerically checked the removal of linear divergence by counterterms 
in DLCQ.
First we consider only $q {\bar q}$ states with instantaneous interaction. 
We study the ground state eigenvalue as a function of $K$
with and without the counterterm. Results are presented in 
Fig. \ref{CT} (a). Next we consider only 
$q {\bar q}$ link states with fermion-link
instantaneous interaction with and without the counterterms. The behavior of 
ground state eigenvalue as a
function of $K$ is presented in Fig. \ref{CT} (b). In both cases, it is
evident that the counterterms are efficient in removing the divergence. 

\subsection{$q {\bar q}$ at the same transverse location}

\begin{figure}
\centering
\includegraphics[width=5.0in]{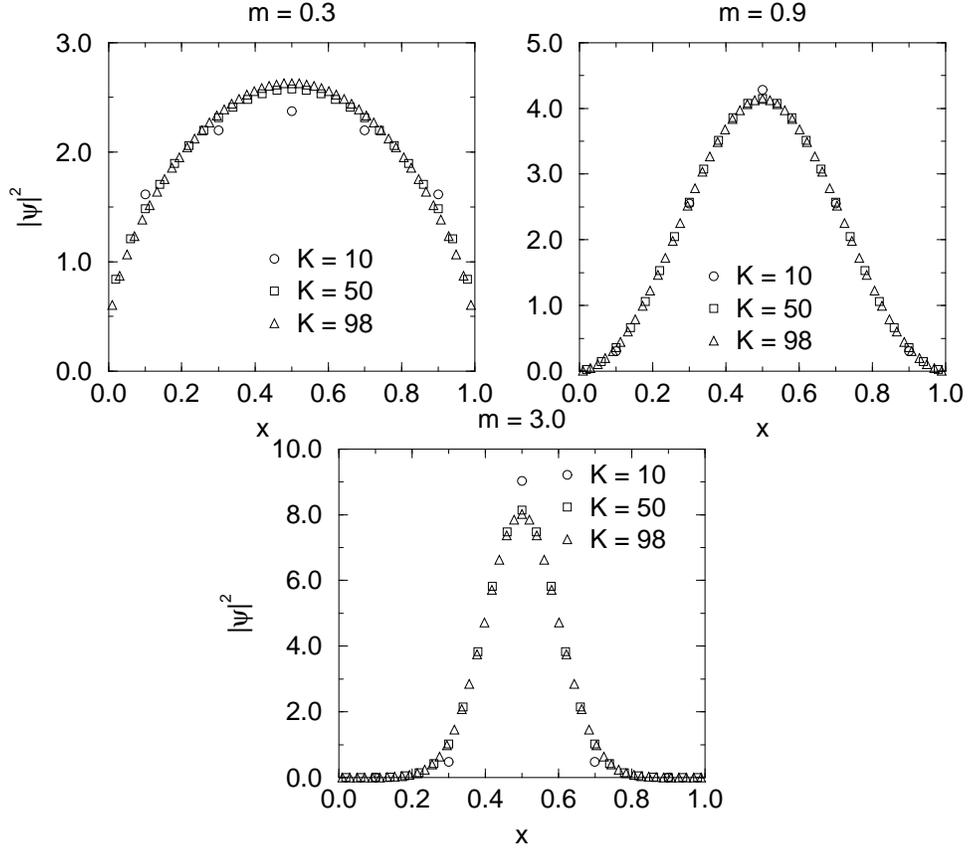}
\caption{Quark distribution function $ \mid \psi(x) \mid^2$ of 
the ground state  in the $q {\bar q}$
approximation for three choices of quark masses with coupling constant $g =1.0$.}
\label{qqbwave}
\end{figure}

Next  we study the spectrum of the Hamiltonian in the absence of any links.
Since, in this case, the Hamiltonian depends
only on the dimensionless ratio $\frac{m_f}{g}$ we fix $g=1$ and vary $m_f$
to study the spectra. 
The Hamiltonian matrix is diagonalized for various
values of $K$. 
The convergence of the ground state eigenvalue as a
function of $K$ is presented
in Table \ref{table1}. 
The ground state wavefunction squared as a function of the
longitudinal momentum fraction $x$ is plotted in Fig.  \ref{qqbwave}. The convergence of
the wavefunction has a very different behavior as a function of fermion 
mass $m_f$. As can be seen from
this figure, the convergence in $K$ is from above for heavy $m_f$   and from
below for light $m_f$. As a consequence the wavefunction  is almost 
independent of $K$
when $m_f$ is of order $g$.     
\subsection{Results of the one link approximation}
\begin{figure}
\centering
\includegraphics[width=5.0in]{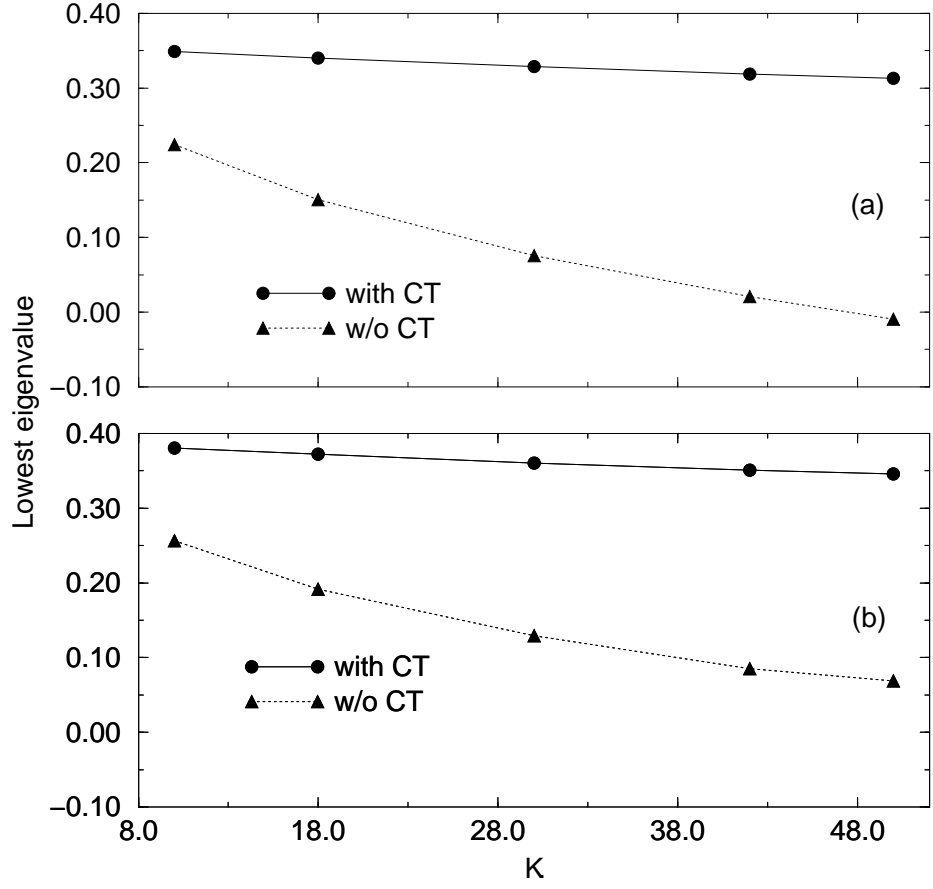}
\caption{Effect of self energy counterterms on the ground state eigenvalue
in the case of (a) symmetric derivative with ${\tilde C}_2=0.4,~ {\tilde C}_3 = 0.1$ 
and (b) forward-backward derivative with $C_2=0.4,~  C_3 = 0.01$.
$m_f =0.3, ~\mu_b=0.2$ for both cases.}
\label{fullself}
\end{figure}

We encountered logarithmic infrared divergences due to self energy
corrections and, in Appendix \ref{appcounter}, we discuss
 the associated
counterterms. In Fig. \ref{fullself} we show the effect of self 
energy counterterms 
on the ground state energy in the two Hamiltonian cases we studied.   

\begin{figure}
\centering
\includegraphics[width=6.0in]{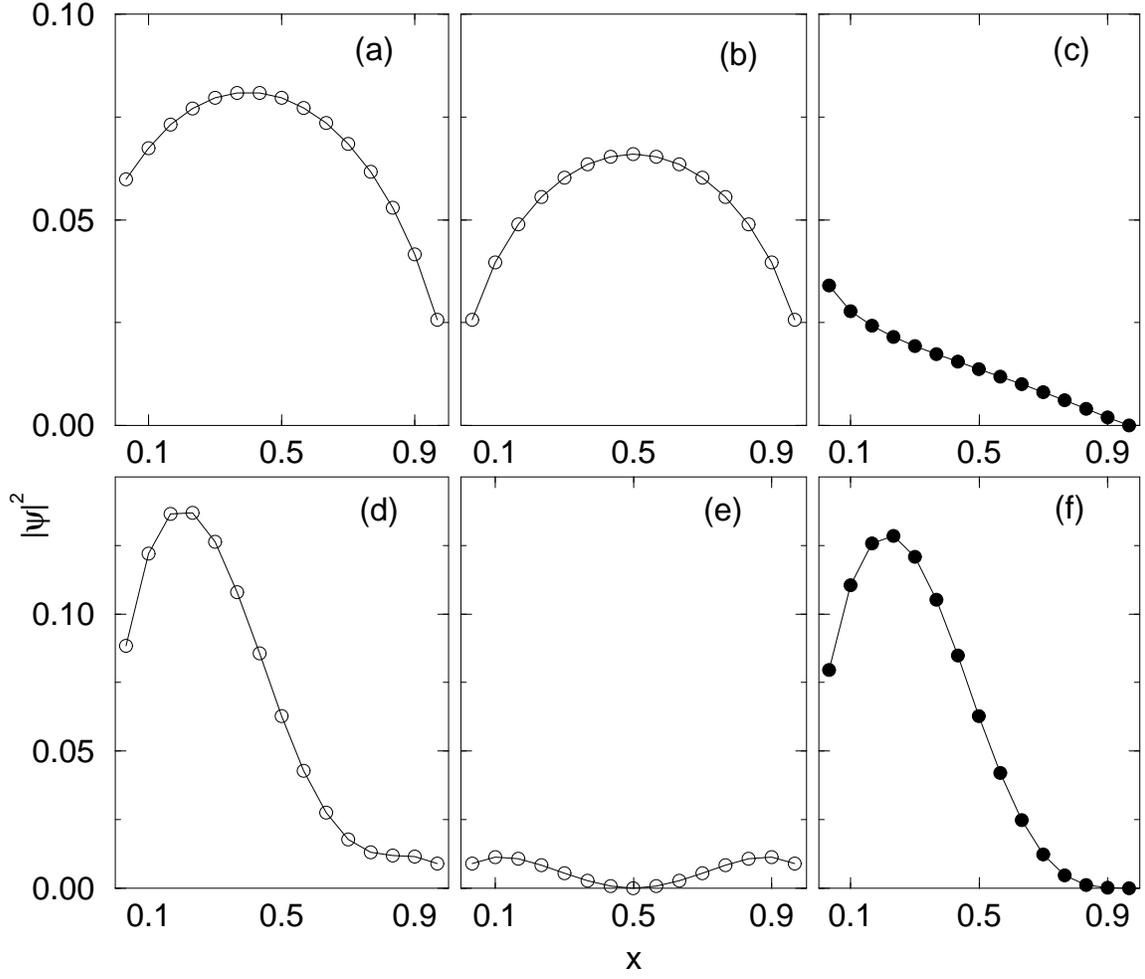}
\caption{ (a) Quark distribution function $ \mid \psi(x) \mid^2$ of the
ground state in the one link
approximation, (b) $q {\bar q}$ contribution to the ground state, 
(c) $q {\bar q}$ link contribution to the ground state. (d)  Quark
distribution function $ \mid \psi(x) \mid^2$ of the
fifth eigenstate in the one link
approximation, (e) $q {\bar q}$ contribution to the fifth eigenstate, 
(f) $q {\bar q}$ link contribution to the fifth eigenstate. The parameters are
$m_f =0.3, ~\mu_b=0.2,  ~C_2=0.4,~  C_3 = 0.01$ and $K = 30$.}
\label{qfull}
\end{figure}

The quark distribution function for the ground state and the fifth state
for the set of parameters  $m_f = 0.3$, $\mu_b = 0.2,~  C_2 = 0.4,
~ C_3 = 0.01$ and $K = 30$ is presented in Fig. \ref{qfull}. 
In this figure we also present separately the contribution from two particle
and three particle states. As expected, the contribution from 
the three particle
state peaks at smaller $x$ compared to the two particle state. The exact
location of this peak depends on the link mass. The convergence of lowest four 
eigenvalues with $K$ for the Hamiltonian with forward-backward and 
symmetric lattice derivatives is shown in Table \ref{table2} for $m_f = 0.3$,
 $\mu_b = 0.2$.  We also show the results extrapolated to $K\rightarrow\infty$.

\begin{figure}
\centering
\includegraphics[width=6.0in]{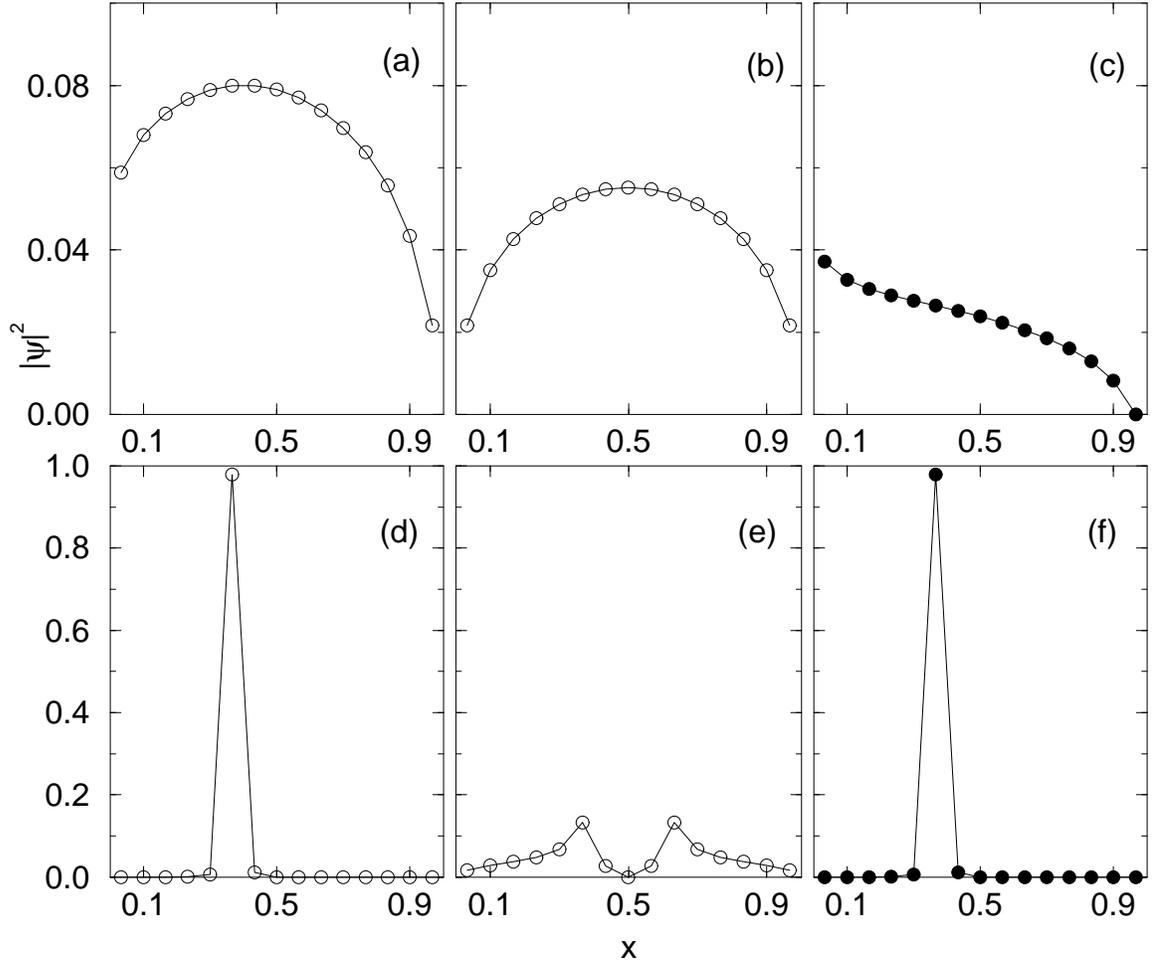}
\caption{ Without the fermion - link instantaneous interaction: 
(a) Quark distribution function $ \mid \psi(x) \mid^2$ of the
ground state in the one link
approximation, (b) $q {\bar q}$ contribution to the ground state, 
(c) $q {\bar q}$ link contribution to the ground state. (d)  Quark
distribution function $ \mid \psi(x) \mid^2$ of the
fifth eigenstate in the one link
approximation, (e) $q {\bar q}$ contribution to the fifth eigenstate
multiplied by $10^{4}$, 
(f) $q {\bar q}$ link contribution to the fifth eigenstate. Parameters are the same as in 
Fig. \ref{qfull}}
\label{q3pfree}
\end{figure}

It is interesting to see the effect of fermion - link instantaneous
interaction on the low lying eigenvalues. In its absence, there is no
confining interaction in the longitudinal direction in the $q {\bar q}$
link sector. Furthermore, the mass of the lowest state  in this sector
corresponds to the threshold mass in this sector. Since its mass is lowered,
it mixes more strongly with the $q {\bar q}$ sector in the ground state. The
fifth state now corresponds to an almost free $q {\bar q}$ link state with
infinitesimal $q {\bar q}$ component as shown in Fig. \ref{q3pfree}.

\section{Summary, Discussion and Conclusions}
In this work we have performed an  investigation  of $q {\bar q}$
states using two different light front Hamiltonians in the one link
approximation. The Hamiltonians correspond to two different ways of
formulating fermions on the transverse lattice, namely, (a) 
forward and backward
derivatives for $\psi^{+}$ and $ \psi^{-}$ respectively or vice versa and (b)
symmetric derivatives for both $\psi^{+}$ and $ \psi^{-}$. In the latter,
fermion doubling is present which is removed by an addition of the Wilson
term. In this case there is no interference between helicity flip hop and
helicity non-flip hop and, as a result, the $q {\bar q}$ component of the
ground state wavefunction which has helicity plus or minus one are
degenerate. In the former case, interference between helicity flip and
helicity non-flip leads to the absence of degeneracy in the low lying spectra. 
One can recover approximate degeneracy of helicity plus or minus one
components only by keeping the strength of the helicity non-flip hopping
very small.
In the case of forward and backward derivatives, terms are also present 
which violate hypercubic
symmetry on the transverse lattice. They become irrelevant in the continuum
limit when  the linear variables $M$ are replaced by non-linear variables $U$. 
 We have removed them
entirely from the Hamiltonian in the present investigation.  

Since the one link approximation is very crude,
we have not attempted a detailed fit to low lying  states in the meson
sector. Instead, we have explored the effects of various coupling strengths
on the low lying spectra and associated wavefunctions. In our work,
longitudinal dynamics is handled by DLCQ. We have performed a detailed study
of various convergence issues in DLCQ  using a wide range of $K$ values.

We summarize our results as follows. We have shown the effectiveness of
appropriate counterterms in the $q {\bar q}$ and $ q {\bar q}$ link sector
to regulate the instantaneous fermion and fermion - link interactions
respectively. We have also checked the cancellation of logarithmic
divergences due to self energy effects. In the limit where fermions are 
frozen on the transverse
lattice but undergo instantaneous longitudinal interaction, we have studied
the convergence of ground state wavefunction with respect to $K$ for three
typical values of the fermion mass. We have studied how the presence or
absence of fermion - link instantaneous interaction in the $q {\bar q}$ link
sector affects the wavefunction of low lying states. We have also studied  
the consequences of the interference of helicity flip and helicity non-flip
hopping in the Hamiltonian with forward-backward derivatives.
This interference is  absent
in the symmetric derivative case.

For future studies, we would like to address the problem of mesons 
containing one light and one heavy quark in the context of heavy quark
effective theory on the transverse lattice. A systematic study of the 
effects of sea quarks and
additional links on the meson observables also 
need to be undertaken.
A major unsettled issue in the transverse lattice formulation is the
continuum limit of the theory when nonlinear link variables are replaced by
link variables. It will be interesting to investigate the light front
quantization problem with non-linear constraints. In this respect 
the study of non-linear sigma model on the light front appears worthwhile.   
 
\acknowledgments
We thank Asit K. De for many
helpful discussions. This work is supported in part by the Indo-US
Collaboration project jointly funded by the U.S. National Science
Foundation (INT0137066) and the Department of Science and Technology, India
(DST/INT/US (NSF-RP075)/2001). This work is also supported in part by the
 US Department of Energy, Grant
No. DE-FG02-87ER40371, Division of Nuclear Physics and a project under
Department of Atomic Energy, India. 

\appendix
\section{Structure of terms in DLCQ}
\label{appa}
We use  DLCQ for the
longitudinal dimension ($ -L \le x^- \le +L$) and  implement
anti periodic boundary condition for  the two component fermion field, 
\be
\eta_c(x^-, \bx) = { 1 \over \sqrt{2L}} \sum_\lambda \chi_\lambda
\sum_{m=1,3,5, \dots} [ b_c(m, \bx, \lambda)e^{-i  \pi m x^- /(2
L)} + d_c^\dagger(m, \bx, -\lambda) e^{i  \pi m x^- /(2 L)}]
\ee
with
\be
\{ b_c (m, \bx, \lambda), b_c^\dagger(m', \bx', \lambda') \} =
 \{ d_c (m, \bx, \lambda), d_c^\dagger(m', \bx', \lambda') \} =
\delta_{m m'} \delta_{\bx, \bx'} \delta_{c,c'} \delta_{\lambda,
\lambda'}.
\ee
The link field has periodic boundary condition (with the omission of the 
zero
momentum mode),
\be
M_{r~pq}(x^-, \bx) = { 1 \over \sqrt{4 \pi}} 
\sum_{m=1,2,3, \dots} \frac{1}{\sqrt{m}}[ B_{-r~pq}(m, \bx + a \br)e^{-i  \pi m x^- / 
L} + B_{r~pq}^\dagger(m, \bx) e^{i  \pi m x^- /L)}]
\ee
with
\be
[ B_{r~pq} (m, \bx), B_{r'~ts}^\dagger(m', \bx') ] =
\delta_{m m'} \delta_{\bx, \bx'} \delta_{r,r'} \delta_{ps} \delta_{qt}.
\ee

The Hamiltonian $P^- = { L \over \pi}H$. 

In the following subsection we give the explicit structure of terms in the
Hamiltonian in the forward-backward case in DLCQ restricting to those 
relevant for the one link approximation. 


\subsection{Mass terms}
Mass terms:

\be
H_{f ~ free} & = &  m^2  \sum_\bx \sum_c\sum_\lambda \sum_n { 1
\over n} \Big [  b_c^\dagger(n, \bx, \lambda) b_c(n, \bx ,\lambda)
+ d_c^\dagger(n, \bx,
\lambda) d_c(n, \bx, \lambda) \Big ].
\ee
\be
H_{LINK ~ free} = \frac{\mu^2}{2} \sum_{\bx} \sum_{\br} \sum_n
\frac{1}{n}
\left [ B_r^\dagger(m,\bx) B_r(m,\bx) + B_{-r}^\dagger(m,\bx + a \br)
B_{-r}(m, \bx + a \br) \right ].
\ee   
%
\subsection{Four fermion  instantaneous term}
The four fermion instantaneous term which gives rise to a linear potential in the
color singlet state
\be
2 { g^2 \over \pi a^2} \sum_{c c'c'' c'''}
\sum_{\lambda \lambda'\lambda'' \lambda'''}  \sum_\bx
\delta_{\lambda \lambda'} \delta_{\lambda'' \lambda'''}
\sum_{m_1 m_2 m_3 m_4} ~~~~~~~~~~~~~~~~~~~~~~~~~~
\nonumber \\
\times ~b^\dagger_c(m_1, \bx, \lambda)
d^\dagger_{c'''}(m_4, \bx, -\lambda''') b_{c'}(m_2, \bx, \lambda')
d_{c''}(m_3, \bx, \lambda''') ~~~~~~~~\nonumber \\
~~~~~~~\times ~{ 1 \over (m_3 -m_4)^2} \delta_{m_1+m_4,
m_2+m_3}~. 
\ee
\subsection{Helicity flip terms}
Particle number conserving terms:

\be
{ m g \over a} \sum_r  \sum_\bx \sum_{\lambda_{1}, \lambda_{2}}
~\chi^\dagger_{\lambda_{1}} ~{\hat \sigma}_r ~\chi_{\lambda_2}~
\sum_{m
_{1}} {1 \over m_{1}} ~~~~~~~ \nonumber \\
\left [ b^\dagger_c(m_{1}, \bx, \lambda_{1}) b_c(m_{1}, \bx, \lambda_{2}) +
d^\dagger_c(m_{1}, \bx, - \lambda_{2}) d_c(m_{1}, \bx, - \lambda_{2})
\right].
\ee

Particle number non conserving terms:
a typical term is
\be
{ m g \over a} { 1 \over \sqrt{4 \pi}}
\sum_r  \sum_\bx \sum_{\lambda_1, \lambda_2}~
~\chi^\dagger_{\lambda_{1}} ~{\hat \sigma}_r ~\chi_{\lambda_2}~
\sum_{m_{1}m_{2}m_{3}} {1 \over \sqrt{m_3}} { 1 \over 2m_3+m_2}
\delta_{m_{1} - m_{2}, 2 m_{3}} \nonumber \\
b_c^\dagger (m_1, \bx, \lambda_1) B_{-r c c '}(m_3, \bx + a \br)
b_{c'}(m_2, \bx + a \br, \lambda_2).
\ee
\subsection{Helicity non flip terms}

Two operators:

\be
{ 2 \over a^2}  \sum_\bx \sum_\lambda \sum_n { 1 \over n}
\left [ b^\dagger_{c}(n,\bx,\lambda) b_{c}(n,\bx, \lambda) +
d^\dagger_{c}(n,\bx,\lambda) d_{c}(n,\bx, \lambda \right ] .
\ee
Three operators:

A typical term is
\be
- g{ 1 \over a^2} { 1 \over \sqrt{4 \pi}} \sum_r  \sum_\bx
\sum_{\lambda}  \sum_{m_1 m_2
m_3} {1 \over \sqrt{m_3}} { 1 \over 2m_3 + m_2} \delta_{m_1 - m_2, 2m_3}
\nonumber \\
b^\dagger_c (m_1, \bx, \lambda) B_{-rcc'}(m_3, \bx + a \br)
b_{c'}(m_2, \bx + a \br, \lambda).
\ee
%
\subsection{Fermion - link instantaneous term}
A typical term is
\be
2 { g^2 \over 4 \pi} { 1 \over a^2}  \sum_\bx \sum_{r} \sum_{cc'c''}
\sum_{dd'} T^\alpha_{cc'} T^\alpha_{dd'} \sum_{m_{1}m_{2}m_{3}m_{4}}
\frac{1}{\sqrt{m_{3}}} \frac{1}{\sqrt{m_{4}}}\nonumber \\
b_d^\dagger(m_1, \bx, \lambda_1) b_{d'}(m_2, \bx, \lambda_2) B_{-r
c'c''}(m_3, \bx + a \br) B^\dagger_{-r c''c}(m_4, \bx + a \br) \nonumber
\\
(-)(m_3+ m_4)/(m_1 - m_2)^2 ~~\delta_{m_{1} - m_{2}, 2m_{3} - 2
m_{4}}~.~~~~~~~~
\ee
\section{States in DLCQ}
\label{appb}
We will consider states of zero transverse momentum.
In the one - link approximation, the gauge invariant states are
 $q {\bar q}$ states
\be  \mid 2 \rangle &= &{ 1 \over \sqrt{N}}~{1 \over \sqrt{V}} ~
\sum_d ~ \sum_{\by(q)}~\sum_{\by({\bar q})} ~ \delta_{\by(q), \by({\bar q})}
\nonumber \\
&~&~~~~b^\dagger_d(n_1, \by(q), \sigma_1)~ 
d^\dagger_d(n_2, \by({\bar q}), \sigma_2)~\mid 0 \rangle 
\ee
and the $q {\bar q} ~{\rm link}$ states 
\be
\mid 3a \rangle &=& { 1 \over N}~{1 \over \sqrt{V}} ~ \frac{1}{\sqrt{2}}
~\sum_{dd'}~ \sum_s~\sum_{\by(q)}~\sum_{\by({\bar q})}~ \sum_{\by(l)}~
\delta_{\by(l),\by(q)}~\delta_{\by(q),\by({\bar q})- a \bs} \nonumber \\
&~&~~~~b^\dagger_{d}(n_1, \by({q}), \sigma_1)~ B^\dagger_{s
dd'}(n_3, \by({l}))~ d^\dagger_{d'}(n_2, \by({\bar q}), \sigma_2)~ 
~\mid 0
\rangle \nonumber 
\ee
and
\be
\mid 3b \rangle &=& { 1 \over N}~{1 \over \sqrt{V}} ~ \frac{1}{\sqrt{2}}
~\sum_{dd'}~ \sum_s~\sum_{\by(q)}~\sum_{\by({\bar q})}~ \sum_{\by(l)}~
\delta_{\by(l),\by(q)}~\delta_{\by(q),\by({\bar q})+ a \bs} \nonumber \\
&~&~~~~b^\dagger_{d}(n_1, \by(q), \sigma_1)~ B^\dagger_{-s
dd'}(n_3, \by(l))~ d^\dagger_{d'}(n_2, \by({\bar q}), \sigma_2)~ 
\mid 0
\rangle . 
\ee
We shall consider transition from these initial states to the following
final states:
The $q {\bar q}$ state
\be
\langle 2' \mid &=& { 1 \over \sqrt{N}}~
~{1 \over \sqrt{V}} ~
\sum_{e}
\sum_{\bz(q)}~\sum_{\bz({\bar q})} ~ \delta_{\bz(q), \bz({\bar q})} 
\nonumber \\
&~&~~~~\langle 0 \mid  
d_e(n_2', \bz({\bar q}), \sigma_2')~
 b_e(n_1', \bz({ q}), \sigma_1')
\ee
and the $q {\bar q} ~{\rm link}$ states 
\be
\langle 3a' \mid &=& { 1 \over N}~ 
{1 \over \sqrt{V}} ~ \frac{1}{\sqrt{2}} \sum_{ee'}~ \sum_t
~ \sum_{\bz(q)}~\sum_{\bz({\bar q})}~ \sum_{\bz(l)}~
\delta_{\bz(l),\bz(q)}~\delta_{\bz(q),\bz({\bar q})- a \bt} \nonumber \\
&~&~~~~\langle 0 \mid  
d_{e}(n_2', \bz({\bar q}), \sigma_2')~ 
B_{te e'}(n_3', \bz(l))~ b_{e'}(n_1', \bz(q), \sigma_1')~
\ee
and
\be
\langle 3b' \mid &=& { 1 \over N} ~{1 \over \sqrt{V}} ~ \frac{1}{\sqrt{2}}
~\sum_{ee'}~
\sum_t
~ \sum_{\bz(q)}~\sum_{\bz({\bar q})}~ \sum_{\bz(l)}~
\delta_{\bz(l),\bz(q)}~\delta_{\bz(q),\bz({\bar q})+ a \bt} \nonumber \\
&~&~~~~\langle 0 \mid  d_{e}(n_2', \bz({\bar q}), \sigma_2')~ 
B_{-te e'}(n_3', \bz(l)) ~b_{e'}(n_1', \bz({q}), \sigma_1')~
 \nonumber \\
\ee 
\section{Forward-backward derivatives: Matrix Elements in DLCQ} 
\label{appc}

\subsection{Transitions from two particle state}
\subsubsection{To two particle state}
Let us consider transitions to the two particle state:
We have, from the free particle term,
\be
\langle 2' \mid H_{f~free} \mid 2 \rangle = m^2  \left ( { 1 \over
n_1} + { 1 \over n_2} \right )~{\cal N}_2
\ee
where 
\be
{\cal N}_2 = \delta_{n_1,n_1'}~
 \delta_{\sigma_1, \sigma_1'}~ 
 \delta_{n_2,n_2'} ~ \delta_{\sigma_2, \sigma_2'}~ .
\ee
From the four fermion instantaneous  term we get
\be
\langle 2' \mid H_{qqc} \mid 2 \rangle &=& - 2 {g^2 \over \pi a^2} 
~C_f   ~ \delta_{n_1 + n_2,n_1' + n_2'}~ 
{ 1 \over (n_1 - n_1')^2 } \nonumber \\
&~&~\delta_{\sigma_1, \sigma_1'} ~ 
\delta_{\sigma_2, \sigma_2'} 
\ee
where $ C_f = {N^2 - 1 \over 2 N}$~.

To implement the regulator prescription for $\frac{1}{(k^+)^2}$, we add the
counterterm matrix elements
\be
\langle 2' \mid H_{CT} \mid 2 \rangle &=&
2 {g^2 \over \pi a^2}~C_f   ~ \delta_{n_1 + n_2,n_1' + n_2'}~
\sum_{n_{loop}=1}^{K} \frac{1}{(n_1 - n_{loop})^2}~
~\delta_{\sigma_1, \sigma_1'} ~\delta_{\sigma_2, \sigma_2'}. 
\ee
Here the term $ n_{loop}=n_1$ is dropped from the sum.

From the helicity flip term we get
\be
\langle 2' \mid H_{hf1} \mid 2 \rangle &=& - 2 { 1 \over a}  \sum_s
\left [ {m  \over n_1} ~\chi^{\dagger}_{\sigma_{1}'}~ 
{\hat \sigma}_s ~\chi_{\sigma_{1}} ~\delta_{\sigma_2, \sigma_2'} ~+
~{m  \over n_2} ~ \chi^\dagger_{-\sigma_{2}} ~
{\hat \sigma}_s ~\chi_{-\sigma_{2}'} ~\delta_{\sigma_1, \sigma_1'} \right ] ~
{\cal N}_{hf} \nonumber \\
\ee
with 
\be
{\cal N}_{hf} = \delta_{n_1,n_1'} 
~\delta_{n_2,n_2'} ~.
\ee
From the helicity non-flip term we get
\be
\langle 2' \mid H_{hnf}(1) \mid 2 \rangle &=&  2 { 1 \over a^2}  
 \left  ( { 1 \over n_1} + { 1 \over n_2} \right ) ~ {\cal N}_2~.
\ee
\subsubsection{To three particle state}
\subsubsection{To the state $\mid 3 a \rangle$}
From the helicity flip term we get
\be
\langle 3 a' \mid H_{hf2} \mid 2 \rangle & = & { mg \over a} 
 ~\sqrt{N}~\frac{1}{V}\frac{1}{\sqrt{2}}~{ 1 \over \sqrt{4 \pi}}~ \sum_t 
\chi^\dagger_{\sigma_1'}~{\hat \sigma}_t ~\chi_{\sigma_1}~
\delta_{\sigma_2, \sigma_2'} \nonumber \\
&~&  \delta_{n_2,n_2'}~{ \delta_{n_1' + 2n_3', 
n_1} \over n_1'}~ {1 \over \sqrt{n_3'}} ~ 
 ~ \sum_{\bz(q)}~ 
\sum_{\by(q)}~ \delta_{\bz(q),\by(q)- a \bt}\nonumber \\
&~&  +~ { mg \over a}  ~\sqrt{N}~\frac{1}{V}\frac{1}{\sqrt{2}}~
{ 1 \over \sqrt{4 \pi}} 
\sum_t~\chi^\dagger_{-\sigma_2}~{\hat \sigma}_t ~\chi_{-\sigma_2'}~
\delta_{\sigma_1, \sigma_1'} \nonumber \\
&~&  \delta_{n_1,n_1'}~{ \delta_{n_2' + 2n_3', 
n_2} \over n_2}~ { 1 \over \sqrt{n_3'}}~
 ~ \sum_{\bz({\bar q})}~ 
\sum_{\by({\bar q})}~ \delta_{\bz({\bar q}),\by({\bar q})+ a \bt}.
\ee 

From the helicity non-flip term we get
\be
\langle 3 a' \mid H_{hnf}(2) 
\mid 2 \rangle & = & -g{ 1 \over a^2}
~\sqrt{N}~\frac{1}{V}~\frac{1}{\sqrt{2}}~ { 1 \over \sqrt{4 \pi}}
~\delta_{\sigma_1, \sigma_1'}
\delta_{\sigma_2, \sigma_2'} \nonumber \\
&~&  \delta_{n_2,n_2'}~{ \delta_{n_1'+ 2n_3',
n_1} \over n_1'}~ {1 \over \sqrt{n_3'}} ~ 
~ \sum_t~\sum_{\bz(q)}~
\sum_{\by(q)}~ \delta_{\bz(q),\by(q)- a \bt}
 \nonumber \\
&~&  -g~ { 1 \over a ^2}  ~\sqrt{N} 
~\frac{1}{V}~\frac{1}{\sqrt{2}}~ { 1 \over \sqrt{4 \pi}}~
\delta_{\sigma_2, \sigma_2'}~
\delta_{\sigma_1, \sigma_1'} \nonumber \\
&~&  \delta_{n_1,n_1'}~{ \delta_{n_2' + 2n_3',n_2} \over n_2}~ 
{ 1 \over \sqrt{n_3'}}
~\sum_t~ \sum_{\bz({\bar q})}~
\sum_{\by({\bar q})}~ \delta_{\bz({\bar q}),\by({\bar q})+ a \bt}~
.
\ee 
\subsubsection{To the state $ \mid 3 b \rangle$}
 
From the helicity flip term we get
\be
\langle 3 b' \mid H_{hf2} \mid 2 \rangle & = & { mg \over a} 
 ~\sqrt{N}~\frac{1}{V}~\frac{1}{\sqrt{2}}~ { 1 \over \sqrt{4 \pi}} ~
~\sum_t~\chi^\dagger_{\sigma_1'}~{\hat \sigma}_t ~\chi_{\sigma_1}~
\delta_{\sigma_2, \sigma_2'} \nonumber \\
&~&  \delta_{n_2, n_2'}~
{ \delta_{n_1' + 2 n_3',
n_1} \over n_1}~ { 1 \over \sqrt{n_3'}} 
~\sum_{\bz({q})}~
\sum_{\by({ q})}~ \delta_{\bz({ q}),\by({q})+ a \bt}~
 \nonumber \\
&~&  +~ { mg \over a}  ~\sqrt{N}
~\frac{1}{V}~\frac{1}{\sqrt{2}}~  {1 \over \sqrt{4 \pi}}~
~\sum_t
\chi^\dagger_{-\sigma_2}~{\hat \sigma}_t ~\chi_{-\sigma_2'}~
\delta_{\sigma_1, \sigma_1'} \nonumber \\
&~&  \delta_{n_1, n_1'}~{ \delta_{n_2' + 2 n_3',
n_2} \over  n_2'}~ {1 \over \sqrt{n_3'}} 
~\sum_{\bz({\bar q})}~
\sum_{\by({\bar q})}~ \delta_{\bz({ \bar q}),\by({\bar q})- a \bt}~
.
\ee 
From helicity non-flip term we get
\be
\langle 3 b' \mid H_{hnf}(3) \mid 2 \rangle 
& = & - g{ 1 \over a^2} ~\sqrt{N}
~\frac{1}{V}~\frac{1}{\sqrt{2}}~  {1 \over \sqrt{4 \pi}}~
\delta_{\sigma_1, \sigma_1'}~
\delta_{\sigma_2, \sigma_2'} \nonumber \\
&~&  \delta_{n_2,n_2'}~
{ \delta_{n_1' + 2 n_3',
n_1} \over n_1}~ { 1 \over \sqrt{n_3'}} ~
\sum_t ~~\sum_{\bz({q})}~
\sum_{\by({ q})}~ \delta_{\bz({ q}),\by({q})+ a \bt}~
 \nonumber \\
&~&  -g~ { 1 \over a^2} ~\sqrt{N}
~\frac{1}{V}~\frac{1}{\sqrt{2}}~  { 1 \over \sqrt{4 \pi}}
\delta_{\sigma_2, \sigma_2'}~
\delta_{\sigma_1, \sigma_1'} \nonumber \\
&~&  \delta_{n_1, n_1'}~
{ \delta_{n_2' + 2 n_3',
n_2} \over  n_2'}~ { 1 \over \sqrt{n_3'}} ~ 
\sum_t~~\sum_{\bz({\bar q})}~
\sum_{\by({\bar q})}~ \delta_{\bz({ \bar q}),\by({\bar q})- a \bt}~
.
\ee 
\subsection{Transitions from three particle ($q~ {\bar q} 
~{\rm link}$) state $ \mid 3 a \rangle $}
\subsubsection{To three particle state}
From the free particle term, we get
\be
\langle 3a' \mid H_{free} \mid 3 a \rangle = \Bigg ( m^2  \Big ( { 1
\over n_1} + { 1 \over n_2} \Big ) + { 1 \over 2} \mu^2  { 1 \over n_3}
\Bigg ) {\cal N}_3
\ee
with 
\be
{\cal N}_3 & = &  \delta_{n_1,n_1'}~
\delta_{n_2, n_2'}~ \delta_{n_3,n_3'} 
 ~\delta_{\sigma_1, \sigma_1'}~ 
\delta_{\sigma_2, \sigma_2'}~.
\ee 

Diagonal contribution from the four  
fermion instantaneous term to the three 
particle state vanishes due to the vanishing trace of 
the generators 
of $SU(N)$.

Contribution from the fermion - link instantaneous term  
\be
\langle 3 a' \mid H_{qgc}(1) \mid 3a \rangle & = & -  {g^2  \over 
\pi}
~{ 1 \over
a^2}  C_f ~ \delta_{n_1+ 2 n_3, n_1'+2 n_3'}~
 \delta_{n_2,n_2'} \nonumber \\
&~&~~~ { 1 \over \sqrt{n_3}\sqrt{n_1 - n_1' + 2 n_3}} ~ 
{ (n_1 - n_1'+4n_3 ) \over (n_1 -n_1')^2}~ \frac{1}{\sqrt{2}}
 ~ \delta_{\sigma_1, \sigma_1'}~ \delta_{\sigma_2,
\sigma_2'} \nonumber \\
&~& -  { g^2 \over  \pi} ~{ 1 \over
a^2}  C_f ~ \delta_{n_2 + 2 n_3, n_2'+2 n_3'}~
 \delta_{n_1,n_1'} \nonumber \\
&~&~~~ { 1 \over \sqrt{n_3}\sqrt{n_2- n_2'+ 2 n_3}} ~ 
{ (n_2 - n_2'+4n_3)
 \over (n_2 -n_2')^2}~ \frac{1}{\sqrt{2}}
 ~ \delta_{\sigma_1, \sigma_1'}~ \delta_{\sigma_2,
\sigma_2'}~. \nonumber \\
\ee
Counterterm matrix elements in DLCQ to implement the regulated prescription
for $\frac{1}{(k^+)^2}$
\be
\langle 3 a' \mid H_{CT}(1) \mid 3a \rangle & = &   {g^2  \over
\pi}
~{ 1 \over
a^2}  C_f ~ \delta_{n_1+ 2 n_3, n_1'+2 n_3'}~
 \delta_{n_2,n_2'}~ \delta_{\sigma_1, \sigma_1'}~ \delta_{\sigma_2,
\sigma_2'} \nonumber \\
&~&~~ \Bigg [ \sum_{n_{loop}=1}^{{n_1}_{max}}{ 1 \over \sqrt{n_3}\sqrt{n_1-
n_{loop}+ 2 n_3}} 
\frac{(n_1 - n_{loop}+4n_3)}{(n_1 - n_{loop})^2}\frac{1}{\sqrt{2}}
\nonumber \\
&~&~~ + \sum_{n_{loop}=1}^{{n_2}_{max}} 
{ 1 \over \sqrt{n_3}\sqrt{n_2-
n_{loop}+ 2 n_3}}
\frac{(n_2 - n_{loop}+4n_3)}{(n_2 - n_{loop})^2} \frac{1}{\sqrt{2}}\Bigg ]~
, \label{flct} \nonumber \\
\ee
where ${n_1}_{max}< n_1+ 2 n_3$ and  ${n_2}_{max}< n_2+ 2 n_3$.
 
The contribution from the helicity flip term that conserves particle number is
\be
\langle 3 a' \mid H_{hf}(1) \mid 3a \rangle & = &
-2 ~{ m \over a}  ~  \delta_{n_1,n_1'}~
 \delta_{n_2,n_2'}~\delta_{n_3, n_3'}  \nonumber \\
&~& ~~~~ \Bigg [ { 1 \over n_1} ~ 
\sum_r ~\chi^\dagger_{\sigma_1'} ~{\hat \sigma_r}~ 
~\chi_{\sigma_{1}} ~ \delta_{\sigma_2, \sigma_2'} ~ +~
{ 1 \over n_2} ~ \sum_r~ \chi^\dagger_{-\sigma_2} {\hat \sigma_r} 
\chi_{-\sigma_{2}'} ~ \delta_{\sigma_1, \sigma_1'} \Bigg ] . \nonumber \\
\ee
The contribution from the helicity non-flip term that conserves particle number
is
\be
\langle 3 a' \mid H_{hnf}(1) \mid 3a \rangle & = & { 2 \over a^2}  
\left ( { 1 \over n_1} + { 1 \over n_2} \right )
{\cal N}_3 ~.
\ee

\subsubsection{To two particle state}
From the helicity flip term we get
\be
\langle 2' \mid H_{hf2} \mid 3a \rangle & = & 
{ mg \over a}  ~\sqrt{N}~ \frac{1}{V}~ \frac{1}{\sqrt{2}}~
{ 1 \over \sqrt{ 4 \pi}} 
\sum_s~\chi^\dagger_{\sigma_1'}~{\hat \sigma}_s ~\chi_{\sigma_1}~
\delta_{\sigma_2, \sigma_2'} \nonumber \\
&~&  \delta_{n_2, n_2'} ~{ \delta_{n_1', 
n_1 +2
n_3} \over n_1}~ { 1 \over \sqrt{n_3}} ~ 
\sum_{\bz({q})}~
\sum_{\by({ q})}~ \delta_{\bz({ q}),\by({q})+ a \bs}~
 \nonumber \\
&~& ~ +~ { mg \over a} ~\sqrt{N}~ 
~ \frac{1}{V}~ \frac{1}{\sqrt{2}}~
{ 1 \over \sqrt{4 \pi}}~
\sum_s~\chi^\dagger_{-\sigma_2}~{\hat \sigma}_s ~\chi_{-\sigma_2'}~
\delta_{\sigma_1, \sigma_1'} \nonumber \\
&~&  \delta_{n_1, n_1'}~{ \delta_{n_2', n_2+ 2n_3 } \over n_2'}~ { 1 \over \sqrt{n_3}} ~ 
\sum_{\bz({\bar q})}~
\sum_{\by({ \bar q})}~ \delta_{\bz({ \bar q}),\by({\bar q}) - a \bs}~
.
\ee 
From the helicity non-flip term we get
\be
\langle 2' \mid H_{hnf} \mid 3a \rangle & = & 
-g{ 1 \over a^2}  ~\sqrt{N}~ \frac{1}{V}~ \frac{1}{\sqrt{2}}~
{ 1 \over \sqrt{4 \pi}}~
\delta_{\sigma_1,\sigma_1'}~
\delta_{\sigma_2, \sigma_2'} \nonumber \\
&~&  \delta_{n_2, n_2'}~{ \delta_{n_1', 
n_1 +
2 n_3} \over n_1}~ { 1 \over \sqrt{n_3}} ~ 
\sum_s \sum_{\bz({q})}~
\sum_{\by({ q})}~ \delta_{\bz({ q}),\by({q})+ a \bs}~
 \nonumber \\
&~&  -g~ { 1 \over a^2}  ~\sqrt{N}~ \frac{1}{V}~ \frac{1}{\sqrt{2}}~
{ 1 \over \sqrt{4 \pi}}~
\delta_{\sigma_2,\sigma_2'}~
\delta_{\sigma_1, \sigma_1'} \nonumber \\
&~&  \delta_{n_1, n_1'}~{ \delta_{n_2', n_2+2n_3 
} \over n_2'}~ { 1 \over \sqrt{n_3}} ~
\sum_{\bz({\bar q})}~
\sum_{\by({ \bar q})}~ \delta_{\bz({ \bar q}),\by({\bar q}) - a \bs}~
.
\ee 

\subsection{Transitions from three particle ($q ~{\bar q}~ 
{\rm link}$) state $ \mid 3 b \rangle $}
\subsubsection{To three particle state}
From the free particle term, we get
\be
\langle 3b' \mid H_{free} \mid 3 b \rangle = \Bigg ( m^2  \Big ( { 1
\over n_1} + { 1 \over n_2} \Big ) + { 1 \over 2} \mu^2  { 1 \over n_3}
\Bigg ) {\cal N}_3
\ee
with 
\be
{\cal N}_3 & = & \delta_{n_1, n_1'}~
\delta_{n_2, n_2'}~ \delta_{n_3, n_3'} 
  ~\delta_{\sigma_1, \sigma_1'}~ 
\delta_{\sigma_2, \sigma_2'}~.
\ee 

The diagonal contribution from the four fermion instantaneous term to the three 
particle state vanishes due to the vanishing trace 
of the generators 
of $SU(N)$.

The contribution from the fermion - link instantaneous term  is
\be
\langle 3 b' \mid H_{qgc}(1) \mid 3b \rangle & = & -  ~{g^2 \over 
\pi}
~{ 1 \over
a^2}  C_f ~ \delta_{n_1 + 2 n_3, n_1' + 2 n_3'}~
 \delta_{n_2, n_2'} \nonumber \\
&~&~~~ 
{ 1 \over \sqrt{n_3}\sqrt{n_1 - n_1' + 2 n_3}} ~
{ (n_1 - n_1'+4n_3 ) \over (n_1 -n_1')^2}~ \frac{1}{\sqrt{2}}
 ~ \delta_{\sigma_1, \sigma_1'}~ \delta_{\sigma_2,
\sigma_2'} \nonumber \\
&~& -  ~{g^2 \over  \pi}~~{ 1 \over
a^2}  C_f ~ \delta_{n_2 +2 n_3, n_2' + 2 n_3'}~
 \delta_{n_1,n_1'} \nonumber \\
&~&~~~ { 1 \over \sqrt{n_3}\sqrt{n_2 - n_2' + 2 n_3}} ~
{ (n_2 - n_2'+4n_3 ) \over (n_2 -n_2')^2}~ \frac{1}{\sqrt{2}}
 ~ \delta_{\sigma_1, \sigma_1'}~ \delta_{\sigma_2,
\sigma_2'}
\ee
Here also we have the counterterm matrix 
elements given in Eq. (\ref{flct}).

The contribution from the helicity flip term that conserves particle number is
\be
\langle 3 b' \mid H_{hf}(1) \mid 3b \rangle & = &
-2 ~{ m \over a}  ~ \delta_{n_1,n_1'}~
 \delta_{n_2, n_2'}~\delta_{n_3, n_3'}  \nonumber \\
&~& ~~~~ \Bigg [ { 1 \over n_1} ~ 
\sum_r ~\chi^\dagger_{\sigma_1'} {\hat \sigma_r} 
~\chi_{\sigma_{1}} ~ \delta_{\sigma_2 \sigma_2'} ~ +~
{ 1 \over n_2} ~ \sum_r~ \chi^\dagger_{-\sigma_2} {\hat \sigma_r} 
\chi_{-\sigma_{2}'} ~ \delta_{\sigma_1 \sigma_1'} \Bigg ] .
\ee
The contribution from the helicity non-flip term that conserves particle number
is
\be
\langle 3 b' \mid H_{hnf}(1) \mid 3b \rangle & = & { 2 \over a^2}  
\left ( { 1 \over n_1} + { 1 \over n_2}  \right )
{\cal N}_3 ~.
\ee
\subsubsection{To the two particle state}
From the helicity flip term we get
\be
\langle 2' \mid H_{hf2} \mid 3b \rangle & = & 
{ mg \over a} ~\sqrt{N}~ \frac{1}{V}~ \frac{1}{\sqrt{2}}~
{ 1 \over \sqrt{4 \pi}}~
\sum_s~ \chi^\dagger_{\sigma_1'}~{\hat \sigma}_s ~\chi_{\sigma_1}~
\delta_{\sigma_2, \sigma_2'} \nonumber \\
&~&  \delta_{n_2, n_2'}~{ \delta_{n_1', 
n_1+ 2 
n_3} \over n_1'}~ { 1 \over \sqrt{n_3}} ~ 
\sum_{\bz({q})}~
\sum_{\by({ q})}~ \delta_{\bz({ q}),\by({q})- a \bs}~
 \nonumber \\
&~&  +~ { mg \over a} ~\sqrt{N}~ 
\frac{1}{V}~ \frac{1}{\sqrt{2}}~
{ 1 \over \sqrt{4 \pi}}~
\sum_s~ \chi^\dagger_{-\sigma_2}~{\hat \sigma}_s ~\chi_{-\sigma_2'}~
\delta_{\sigma_1, \sigma_1'} \nonumber \\
&~&  \delta_{n_1, n_1'}~{ \delta_{n_2', 2 n_3 +
n_2} \over n_2}~ { 1 \over \sqrt{n_3}} ~ 
\sum_{\bz({\bar q})}~
\sum_{\by({ \bar q})}~ \delta_{\bz({ \bar q}),\by({\bar q}) + a \bs}~
~.
\ee 
From the helicity non-flip term we get
\be
\langle 2' \mid H_{hnf} \mid 3b \rangle & = & 
-g{ 1 \over a^2} ~\sqrt{N}~ \frac{1}{V}~ \frac{1}{\sqrt{2}}
{ 1 \over \sqrt{4 \pi}}~
\delta_{\sigma_1,\sigma_1'}~
\delta_{\sigma_2, \sigma_2'} \nonumber \\
&~&  \delta_{n_2, n_2'}~{ \delta_{n_1', 
n_1 + 2
n_3} \over n_1'}~ { 1 \over \sqrt{ n_3}} ~ 
\sum_s~\sum_{\bz({q})}~
 \sum_{\by({ q})}~ \delta_{\bz({ q}),\by({q})- a \bs}~
 \nonumber \\
&~&  -g~ { 1 \over a^2} ~\sqrt{N}~ \frac{1}{V}~ \frac{1}{\sqrt{2}}
{ 1 \over \sqrt{4 \pi}}~
\delta_{\sigma_2,\sigma_2'}~
\delta_{\sigma_1, \sigma_1'} \nonumber \\
&~&  \delta_{n_1, n_1'}~{ \delta_{n_2',  n_2 + 2 n_3 
} \over n_2}~ { 1 \over \sqrt{n_3}} ~ 
\sum_s~\sum_{\bz({\bar q})}~
 ~ \sum_{\by({ \bar q})}~ \delta_{\bz({ \bar q}),\by({q})+ a \bs}~ 
~.
\ee 
\section{Symmetric derivatives and Wilson term: Matrix elements in DLCQ}
\label{appd}
In this section, we list only those matrix elements that differ from the
forward-backward case.
\subsection{Transitions from the two particle state}
\subsubsection{To the state $ \mid 3 a \rangle$}
{ Helicity flip}:
\be
\langle  3a' \mid  P^-_{whf} \mid 2  \rangle & = &
\left (m + 4 \frac{\kappa}{a} \right) \frac{1}{2a} ~ \sqrt{N} 
~\frac{1}{V}~\frac{1}{{\sqrt{2}}}~{ 1 \over
\sqrt{4 \pi}} 
\sum_t ~ \chi^\dagger_{\sigma_{1}'}~
{\hat \sigma}_t ~\chi_{\sigma_{1}} ~\delta_{\sigma_{2}, \sigma_{2}^{'}}
 \nonumber \\
&~& \sum_{\by(q)}~
\sum_{\bz(q)} ~
 \delta_{\bz(q), \by(q)-a \bt}  \nonumber \\
&~&~~~ { 1 \over \sqrt{n_3^{'}}}\left 
( { 1 \over n_1} - {1 \over n_1^{'}} \right )~
 \delta_{n_2,n_2^{'}} ~ \delta_{n_1^{'} + 2 n_3^{'},
n_1} \nonumber \\
&+& \left(m + 4 \frac{\kappa}{a}\right)\frac{1}{2a} ~ \sqrt{N} 
~\frac{1}{V}~\frac{1}{{\sqrt{2}}}
{ 1 \over
\sqrt{4
\pi}}~
\sum_t~\chi^\dagger_{-\sigma_{2}}~
{\hat \sigma}_t ~\chi_{-\sigma_{2}'}~ \delta_{\sigma_{1}, 
\sigma_{1}^{'}}
 \nonumber \\
&~& \sum_{\by({\bar q})} 
~\sum_{\bz({\bar q})} 
~ \delta_{\bz({\bar q}), \by({\bar q})+ a \bt} 
\nonumber \\
&~&~~~ { 1 \over \sqrt{n_3^{'}}}
\left ( { 1 \over n_2^{'}} - {1 \over n_2^{}} \right )~
 \delta_{n_1,n_1^{'}}~  \delta_{n_2^{'} + 2 n_3^{'},
n_2}~. 
\ee
{ Helicity non-flip}:

\be
\langle 3a' \mid  P^-_{wnf1} \mid 2 \rangle & = &
 -\left(m + 4 \frac{\kappa}{a} \right) \frac{\kappa}{a} ~ \sqrt{N} 
~\frac{1}{V}~\frac{1}{{\sqrt{2}}}
{ 1 \over \sqrt{4 \pi}}
 \delta_{\sigma_{2}, \sigma_{2}^{'}}~\delta_{\sigma_{1}, 
\sigma_{1}^{'}}
 \nonumber \\
&~& \sum_t~ \sum_{\by(q)}~
\sum_{\bz(q)} 
\delta_{\bz(q), \by(q)-a \bt}  \nonumber \\
&~&~~~ { 1 \over \sqrt{n_3^{'}}}\left ( { 1 \over n_1} + {1 \over
n_1^{'}} \right )~
 \delta_{n_2,n_2^{'}} ~ \delta_{n_1^{'} + 2 n_3^{'},
n_1} \nonumber \\
&-& \left(m + 4 \frac{\kappa}{a} \right)\frac{\kappa}{a} ~ \sqrt{N} 
~\frac{1}{V}~\frac{1}{{\sqrt{2}}}
{ 1 \over \sqrt{4 \pi}}
 \delta_{\sigma_{2}, \sigma_{2}^{'}~}\delta_{\sigma_{1}, 
\sigma_{1}^{'}}
 \nonumber \\
&~& \sum_t~ \sum_{\by({\bar q})}~
~ \sum_{\bz({\bar q})} 
~~~\delta_{\bz({\bar q}), \by({\bar q})+a \bt} 
\nonumber \\
&~&~~~ { 1 \over \sqrt{n_3^{'}}}\left ( { 1 \over n_2^{'}} + {1 \over
n_2^{}} \right )~
 \delta_{n_1,n_1^{'}}~  \delta_{n_2^{'} + 2n_3^{'},
n_2}~. 
\ee
\subsubsection{To the state $ \mid 3 b \rangle $}
{ Helicity flip}:
 \be
 \langle 3b' \mid  P^-_{whf} \mid 2  \rangle
&=& \left(m + 4 \frac{\kappa}{a}\right) \frac{1}{2a} ~ \sqrt{N} 
\frac{1}{V} \frac{1}{\sqrt{2}}
{1 \over \sqrt{4 \pi}}~
\sum_t ~\chi^\dagger_{\sigma_{1}'}~
{\hat \sigma}_t ~\chi_{\sigma_{1}} \delta_{\sigma_{2}, \sigma_{2}^{'}}
 \nonumber \\
&~& \sum_{\by(q)}~
\sum_{\bz(q)}~ 
~~~  \delta_{\bz(q), \by(q)+a \bt} 
\nonumber \\
&~&~~~ { 1 \over \sqrt{n_3^{'}}}\left (- { 1 \over n_1} + {1 \over
n_1^{'}} \right )~
 \delta_{n_2,n_2^{'}}  ~\delta_{n_1^{'} + 2 n_3^{'},
n_1}~ \nonumber \\
&+& \left(m + 4 \frac{\kappa}{a}\right)\frac{1}{2a} ~ \sqrt{N} 
\frac{1}{V} \frac{1}{\sqrt{2}}
{ 1 \over \sqrt{4 \pi}} 
\sum_t~ \chi^\dagger_{-\sigma_{2}}~
{\hat \sigma}_t ~\chi_{-\sigma_{2}'} ~\delta_{\sigma_{1}, 
\sigma_{1}^{'}}
 \nonumber \\
&~& \sum_{\by({\bar q}}~ \sum_{\bz({\bar q})}
~~~  \delta_{\bz({\bar q}), \by({\bar q}) -a \bt} 
\nonumber \\
&~&~~~ { 1 \over \sqrt{n_3^{'}}}
  \left ( -{ 1 \over n_2^{'}} + {1 \over n_2} \right )~
 ~\delta_{n_1, n_1^{'}} ~ \delta_{n_2^{} + 2 n_3^{},
n_2^{'}}~. 
\ee
{ Helicity non-flip}:
\be
 \langle 3a' \mid P^-_{wnf1} \mid 2  \rangle 
&=&-\left(m + 4 \frac{\kappa}{a}\right) \frac{\kappa}{a} ~ \sqrt{N} 
\frac{1}{V} \frac{1}{\sqrt{2}}
{ 1 \over \sqrt{4 \pi}}
 \delta_{\sigma_{2}, \sigma_{2}^{'}} ~\delta_{\sigma_{1}, \sigma_{1}^{'}}
 \nonumber \\
&~& \sum_t~\sum_{\by(q)}~
~\sum_{\bz(q)}~ 
~~~  \delta_{\bz(q), \by(q)+a \bt} 
\nonumber \\
&~&~~~ { 1 \over \sqrt{n_3^{'}}}
 \left ( { 1 \over n_1} + {1 \over n_1^{'}} 
\right )~
 \delta_{n_2, n_2^{'}} ~\delta_{n_1^{'} + 2 n_3^{'},
n_1}~
 \nonumber \\
&-& \left(m + 4 \frac{\kappa}{a}\right)\frac{\kappa}{a} ~ \sqrt{N} 
\frac{1}{V} \frac{1}{\sqrt{2}}~
{ 1 \over \sqrt{4 \pi}}
 \delta_{\sigma_{1}, \sigma_{1}^{'}} ~\delta_{\sigma_{2}, \sigma_{2}^{'}}
 \nonumber \\
&~& \sum_t ~ \sum_{\by({\bar q})}
~ \sum_{\bz({\bar q})}  
~~~ \delta_{\bz({\bar q}), \by({\bar q}) - a \bt} 
\nonumber \\
&~&~~~ { 1 \over \sqrt{n_3^{'}}}     \left ( { 1 \over n_2^{'}} + {1 \over
n_2} \right )~
~ \delta_{n_1,n_1^{'}} ~ \delta_{n_2^{'} + 2 n_3^{'},
n_2}~. 
\ee

\subsection{Transitions from three particle state $\mid 3 a \rangle $ to two
particle state}
{ Helicity flip}:
\be
\langle 2' \mid P^-_{whf} \mid 3a  \rangle
&=&\left(m + 4 \frac{\kappa}{a} \right) \frac{1}{2a}  \sqrt{N} 
\frac{1}{V} \frac{1}{\sqrt{2}}~
\frac{1}{\sqrt{4 \pi}}
\sum_s~ \chi^\dagger_{\sigma_{1}'}~
{\hat \sigma}_s ~\chi_{\sigma_{1}}~ \delta_{\sigma_{2}, \sigma_{2}^{'}}
 \nonumber \\
&~& \sum_{\bz(q)}
\sum_{\bz(q)} 
~~~ \delta_{\bz(q),\by(q) + a \bs)} 
\nonumber \\
&~&~~~ { 1 \over \sqrt{n_3^{}}} \left ( { 1 \over n_1^{'}} - {1 \over
n_1^{}} \right )~
 \delta_{n_2,n_2^{'}}~  \delta_{n_1 + 2 n_3,
n_1^{'}}~ \nonumber \\
& + & \left(m + 4 \frac{\kappa}{a}\right)\frac{1}{2a} \sqrt{N} 
\frac{1}{V} \frac{1}{\sqrt{2}}~
{1 \over \sqrt{4 \pi}}
\sum_s ~ \chi^\dagger_{-\sigma_{2}}~
{\hat \sigma}_s ~\chi_{-\sigma_{2}'}~ \delta_{\sigma_{1}, \sigma_{1}^{'}}
 \nonumber \\
&~& \sum_{\bz({\bar q})}
\sum_{\by({\bar q})} 
~~ \delta_{\bz({\bar q}),
\by({\bar q}) - a \bs)} 
\nonumber \\
&~&~~~ { 1 \over \sqrt{n_3}}\left ( { 1 \over n_2} - {1 \over
n_2^{'}} \right )~
 \delta_{n_1^{'}, n_1} ~ \delta_{n_2^{'} + 2 n_3^{'},
n_2}~. 
\ee

{ Helicity non-flip}:

\be
\langle 2' \mid  P^-_{wnf1} \mid 3a \rangle
&=& -\left(m + 4 \frac{\kappa}{a}\right) \frac{\kappa}{a}  \sqrt{N} 
\frac{1}{V} \frac{1}{\sqrt{2}}~
{ 1 \over \sqrt{4 \pi}}
 \delta_{\sigma_2, \sigma_{2}^{'}} ~\delta_{\sigma_1, \sigma_{1}^{'}}
 \nonumber \\
&~& \sum_s~ \sum_{\bz(q)}
\sum_{\by(q)}
~~~  \delta_{\bz(q),\by(q) + a \bs)} 
\nonumber \\
&~&~~~ { 1 \over \sqrt{n_3}}\left ( { 1 \over n_1} + {1 \over
n_1^{'}} \right )~
 \delta_{n_2,n_2^{'}} ~ \delta_{n_1 + 2 n_3,
n_1^{'}}~ \nonumber \\
&-& \left(m + 4 \frac{\kappa}{a}\right)\frac{\kappa}{a}  \sqrt{N} 
\frac{1}{V} \frac{1}{\sqrt{2}}~
{ 1 \over \sqrt{4 \pi}} 
 \delta_{\sigma_2, \sigma_{2}^{'}}\delta_{\sigma_1, \sigma_{1}^{'}}
\nonumber \\
&~& \sum_{\bz({\bar q})} ~
\sum_{\by({\bar q})} ~
~\delta_{\bz({\bar q}),\by({\bar q}) + a \bs)} 
\nonumber \\
&~&~~~ { 1 \over \sqrt{n_3}}\left ( { 1 \over n_2^{'}} + {1 \over
n_2} \right )~
 ~\delta_{n_1,n_1^{'}} ~ \delta_{n_2^{} + 2 n_3^{},
n_2^{'}}~.
\ee

\subsection{Transitions from three particle state $\mid 3 b \rangle $ to two 
particle state}

{ Helicity flip}:
\be
\langle 2' \mid P^-_{whf} \mid 3b \rangle
 &=&\left(m + 4 \frac{\kappa}{a}\right) \frac{1}{2a}  \sqrt{N} 
\frac{1}{V} \frac{1}{\sqrt{2}}~
{ 1 \over \sqrt {4 \pi}}
\sum_s ~ \chi^\dagger_{\sigma_{1}'}~
{\hat \sigma}_s ~\chi_{\sigma_{1}}~ \delta_{\sigma_2, \sigma_{2}^{'}}
 \nonumber \\
&~& \sum_{\bz(q)}~
\sum_{\by(q)}~ 
~ \delta_{\bz(q),\by(q) - a \bs)}
\nonumber \\
&~&~~~ { 1 \over \sqrt{n_3}}\left ( { 1 \over n_1} - {1 \over
n_1^{'}} \right )~
 \delta_{n_2,n_2^{'}}~  \delta_{n_1 + 2 n_3,
n_1^{'}}~ \nonumber \\
&+& \left(m + 4 \frac{\kappa}{a}\right)\frac{1}{2a}  \sqrt{N} 
\frac{1}{V} \frac{1}{\sqrt{2}}~
{ 1 \over \sqrt{4 \pi}}
\sum_s ~ \chi^\dagger_{-\sigma_{2}}~
{\hat \sigma}_s ~\chi_{-\sigma_{2}'}~ \delta_{\sigma_1, \sigma_{1}^{'}}
 \nonumber \\
&~& \sum_{\bz({\bar q})} 
\sum_{\by({\bar q})}  
~\delta_{\bz({\bar q}),\by({\bar q}) + a \bs)}
\nonumber \\
&~&~~~ { 1 \over \sqrt{n_3}}\left ( { 1 \over n_2^{'}} - {1 \over
n_2} \right )~
 \delta_{n_1,n_1^{'}}~  \delta_{n_2^{} + 2 n_3^{},
n_2^{'}}~. 
\ee

{ Helicity non-flip}:
\be
\langle 2' \mid  P^-_{whf} \mid 3b \rangle
&=&-\left(m + 4 \frac{\kappa}{a}\right) \frac{\kappa}{a}  \sqrt{N} 
\frac{1}{V} \frac{1}{\sqrt{2}}~
{ 1 \over \sqrt{4 \pi}} 
 \delta_{\sigma_2, \sigma_{2}^{'}}~\delta_{\sigma_1, \sigma_{1}^{'}}
 \nonumber \\
&~& \sum_s~ \sum_{\bz(q)} 
~\sum_{\by(q)}~
 \delta_{\bz(q),\by(q) - a \bs)} 
\nonumber \\
&~&~~~ { 1 \over \sqrt{n_3}}\left ( { 1 \over n_1} + {1 \over
n_1^{'}} \right )~
 \delta_{n_2,n_2^{'}} ~ \delta_{n_1 + 2 n_3,
n_1^{'}}~ \nonumber \\
&-& \left(m + 4 \frac{\kappa}{a}\right)\frac{\kappa}{a}  \sqrt{N} 
\frac{1}{V} \frac{1}{\sqrt{2}}~
{ 1 \over \sqrt{4 \pi}}
 \delta_{\sigma_2, \sigma_{2}^{'}}~\delta_{\sigma_1, \sigma_{1}^{'}}
 \nonumber \\
&~& \sum_s~ \sum_{\bz({\bar q})}
~\sum_{\by({\bar q})}~~ 
~\delta_{\bz({\bar q}),\by({\bar q}) + a \bs)} 
\nonumber \\
&~&~~~ { 1 \over \sqrt{n_3}}\left ( { 1 \over n_2^{'}} + {1 \over
n_2} \right )~
 \delta_{n_1,n_1^{'}} ~ \delta_{n_2^{} + 2 n_3^{},
n_2^{'}}~. 
\ee 
\section{Self energy counterterms}
\label{appcounter}
In this Appendix we list the self energy counterterms. 
\subsection{Symmetric derivatives case}
The counterterm for self energy for a quark or an antiquark with longitudinal
momentum $n_1$ due to double helicity flip hops
\be
CT_1 = \frac{2}{n_1}\sum_{n_{1}'=1}^{n_{1}} \frac{1}{n_{1}'}
\frac{(n_1 - n_1')^2}{\mu^2 n_1 n_1' + m^2 (n_1 - n_1')^2}.
\ee
The counterterm for self energy for a quark or an antiquark with longitudinal 
momentum $n_1$ due to double helicity non-flip hops
\be
CT_2 = \frac{2}{n_1}\sum_{n_{1}'=1}^{n_{1}} \frac{1}{n_{1}'}
\frac{(n_1 + n_1')^2}{\mu^2 n_1 n_1' + m^2 (n_1 - n_1')^2}.
\ee

\subsection{Forward and backward derivative case}
 In this case we have three types of contributions: (1) helicity flip acting
twice, (2) helicity non-flip acting twice and (3) interference of helicity
flip and helicity non-flip hops. The first two are diagonal in helicity
space but the last one is off-diagonal in helicity space.

The transition from state $ \mid 2  \rangle $ to state $ \mid 3 a \rangle $ and
back due to a quark hop gives rise to longitudinal infrared divergence. In
this case the
counterterm due to double helicity flip is
\be 
CT_3 = 2 \sum_{n_{1}'=1}^{n_{1}} \frac{1}{n_{1}'}
\frac{n_1}{\mu^2 n_1 n_1' + m^2 (n_1 - n_1')^2}. \label{ctq}
\ee
The counterterm due to double helicity non-flip is the same without the factor
of 2. The transition from state $ \mid 2  \rangle $ to state $ \mid 3 b \rangle $ and 
back due to a quark hop does not give rise to longitudinal infrared
divergence.
Similarly the transition from state $ \mid 2  \rangle $ to state $ \mid 3 a
\rangle $ and 
back due to an antiquark hop does not give rise to longitudinal infrared
divergence.
The transition from state $ \mid 2  \rangle $ to state $ \mid 3 b \rangle $ and 
back due to an antiquark hop gives rise to longitudinal infrared divergence
which requires  counterterms the explicit forms of which are the same as in
the quark case for the transition from  $ \mid 2  \rangle $ to state $ \mid 3
a \rangle $.
Lastly we consider counterterms for self energy contributions arising from
the interference of helicity flip and helicity non-flip hopping. The
counterterms have the same structure as in the case of helicity non-flip
transitions accompanied by the following extra factors.
Since we have two possibilities namely helicity flip followed by helicity
non-flip and vice versa and these two contributions are the same, we get a
factor of two. We also get a  factor 
$ \chi^\dagger_{s'} {\hat \sigma^\perp} \chi_{s}$
where $s (s')$ is the initial (final) helicity and ${\hat \sigma^1} =
\sigma^2$, ${\hat \sigma^2} =
- \sigma^1$. 



\eject
\begin{table}
\begin{center}
\begin{tabular}{||c|c|c|c||}
\hline \hline
  K & \multicolumn{3}{c||}  {Eigenvalue (${\cal M}^2$)} \\
\cline{2-4}
  & $m_f=0.3$ & $m_f = 0.9$ & $m_f = 3.0$\\
\hline
   10 & 0.620 & 4.547 & 39.233\\
  18 &   0.693 & 4.664 & 39.861\\ 
  30  &   0.745 & 4.724 & 40.053 \\
  50 &   0.788  & 4.762 &  40.163\\
  78 &   0.819 &  4.783  &  40.220\\
  98 &   0.832 & 4.791 &  40.241\\
$K\rightarrow\infty$ & 0.869 & 4.820 & 40.285 \\
\hline \hline
\end{tabular}
\end{center}
\caption{Ground state eigenvalue (in units of $G^2$) for $q {\bar q}$ sitting at 
the same transverse location.}
\label{table1}
\end{table}

\begin{table}
\begin{center}
\begin{tabular}{||c|c|c|c|c||c|c|c|c||}
\hline \hline
  &\multicolumn{4}{c||} {Forward-backward } & 
\multicolumn{4}{c||} {Symmetric }\\
 & \multicolumn{4}{c||} {($ C_2=0.01,~ C_3=0.4$)} & 
\multicolumn{4}{c||} { (${\tilde C}_2=0.1,~{\tilde C}_3=0.4$)}\\
\cline{2-9}
 K & ${\cal M}_1^2$ & ${\cal M}_2^2$ &  ${\cal M}_3^2$ & ${\cal M}_4^2$  
& ${\cal M}_1^2$ & ${\cal M}_2^2$ &  ${\cal M}_3^2$ & ${\cal M}_4^2$\\
\hline
10 & 0.38041 & 0.4800 & 0.4899 & 0.5996 & 0.3486 & 0.4507 & 0.4507 & 0.5980 \\
 18 & 0.3722 & 0.4968 &  0.5110&  0.6447 & 0.3402 & 0.4673 & 0.4673 & 0.6409  \\
 30 & 0.3606 & 0.5027 & 0.5210 &  0.6680& 0.3288 & 0.4702 & 0.4702 & 0.6620 \\
42  & 0.3511  & 0.5029 & 0.5240 & 0.6765&0.3189 & 0.4677 & 0.4677 & 0.6682 \\
50  &  0.3457 & 0.5019 & 0.5246 & 0.6790&0.3130 & 0.4651 & 0.4651 & 0.6693\\
$K\rightarrow\infty$ & 0.3243 & 0.5022 & 0.5313 & 0.6979 & 0.2913 & 0.4589 &
0.4589 & 0.6837 \\
\hline \hline
\end{tabular}
\end{center}
\caption{Lowest four eigenvalues (in units of $G^2$) in one link approximation.}
\label{table2}
\end{table}

\end{document}